\documentclass[myepj-spec]{mySvjour}
\usepackage{graphicx}
\usepackage{hyperref}
\usepackage{color}
\usepackage{xspace}
\usepackage[square,sort&compress,numbers]{natbib}   
\usepackage{verbatim}
\usepackage{amsmath,amssymb,amsfonts} 
\usepackage{tabularx}

\usepackage{lmodern,bm}                
\usepackage[T1]{sansmath} 
\SetMathAlphabet{\mathsfbf}{sans}{\sansmathencoding}{\sfdefault}{bx}{sl}
\usepackage{etoolbox}
\AtBeginEnvironment{sansmath}{}{}{}

\newcommand{\figsize}{\columnwidth}
\newcommand{\fighspace}{0.4cm}

\definecolor{darkblue1}{rgb}{0,0,.2}
\definecolor{darkblue}{rgb}{0,0,.2}
\definecolor{darkred}{rgb}{0.5,0,0}
\pagecolor{white} 
\color{black}     
\hypersetup{breaklinks=true, 
            colorlinks=true, 
            linkcolor=darkblue1, 
            menucolor=darkblue1, 
            urlcolor=darkblue1,
            citecolor=darkblue1,
            pdftitle={},
            pdfauthor={},
            pdfsubject={},
            pdfkeywords={},
            pdfproducer={}
}
\bibstyle{plain}
\mathchardef\Upsilon="7107
\def\Y#1S{\ensuremath{\Upsilon{(#1S)}}\xspace}

\newcommand{\mZ}{\ensuremath{m_Z}\xspace}
\newcommand{\as}{\ensuremath{\alpha_{\scriptscriptstyle S}}\xspace}

\newcommand{\asZ}{\ensuremath{\as(\mZ^2)}\xspace}

\newcommand{\aZ}{\ensuremath{\alpha(m_Z^2)}\xspace}
\newcommand{\dahadZ}{\ensuremath{\Delta\alpha_{\rm had}(m_Z^2)}\xspace}

\newcommand\chiS{\ensuremath{\chi^2}\xspace}
\newcommand\chiSdof{\ensuremath{\chi^2/{\rm ndof}}\xspace}
\newcommand\sqrtS{\ensuremath{\sqrt{\rm s}}\xspace}

\newcommand{\Kbar    }{\kern 0.2em\overline{\kern -0.2em K}{}\xspace}
\newcommand{\Dbar    }{\kern 0.2em\overline{\kern -0.2em D}{}\xspace}
\newcommand{\pbar    }{\kern 0.1em\overline{\kern -0.1em p}{}\xspace}
\newcommand{\nbar    }{\kern 0.1em\overline{\kern -0.1em n}{}\xspace}

\newcommand{\Kz      }{\ensuremath{K^0}\xspace}
\newcommand{\Kzb     }{\ensuremath{\Kbar^0}\xspace}
\newcommand{\KzKzb   }{\ensuremath{\Kz \kern -0.16em \Kzb}\xspace}
\newcommand{\Kp      }{\ensuremath{K^+}\xspace}
\newcommand{\Km      }{\ensuremath{K^-}\xspace}

\newcommand{\KpKm    }{\ensuremath{\Kp \kern -0.16em \Km}\xspace}

\newcommand{\pip}{\ensuremath{\pi^+}\xspace}
\newcommand{\pim}{\ensuremath{\pi^-}\xspace}
\newcommand{\piz}{\ensuremath{\pi^0}\xspace}

\newcommand{\ee}{\ensuremath{e^+e^-}\xspace}

\newcommand{\pp}{\ensuremath{\pi^+\pi^-}\xspace}
\newcommand{\pipi}{\ensuremath{\pi^+\pi^-}\xspace}

\newcommand{\ppg}{\ensuremath{\pi^+\pi^-(\gamma)}\xspace}

\newcommand{\tev}{\ensuremath{\mathrm{\,Te\kern -0.1em V}}\xspace}
\newcommand{\gev}{\ensuremath{\mathrm{\,Ge\kern -0.1em V}}\xspace}
\newcommand{\mev}{\ensuremath{\mathrm{\,Me\kern -0.1em V}}\xspace}
\newcommand{\kev}{\ensuremath{\mathrm{\,ke\kern -0.1em V}}\xspace}
\newcommand{\ev}{\ensuremath{\mathrm{\,e\kern -0.1em V}}\xspace}
\newcommand{\gevc}{\ensuremath{{\mathrm{\,Ge\kern -0.1em V\!/}c}}\xspace}
\newcommand{\mevc}{\ensuremath{{\mathrm{\,Me\kern -0.1em V\!/}c}}\xspace}
\newcommand{\gevcc}{\ensuremath{{\mathrm{\,Ge\kern -0.1em V\!/}c^2}}\xspace}
\newcommand{\mevcc}{\ensuremath{{\mathrm{\,Me\kern -0.1em V\!/}c^2}}\xspace}

\newcommand{\bei}{\begin{itemize}}
\newcommand{\eei}{\end{itemize}}
\newcommand{\ben}{\begin{enumerate}}
\newcommand{\een}{\end{enumerate}}
\newcommand{\beq}{\begin{equation}}
\newcommand{\eeq}{\end{equation}}
\newcommand{\beqn}{\begin{eqnarray}}
\newcommand{\eeqn}{\end{eqnarray}}
\newcommand{\beqns}{\begin{eqnarray*}}
\newcommand{\eeqns}{\end{eqnarray*}}

\newcommand{\amu}{\ensuremath{a_\mu}\xspace}

\newcommand{\amuhadLO}{\ensuremath{\amu^{\rm had,LO}}\xspace}
\newcommand{\amuhadLOpp}{\ensuremath{\amu^{\rm had,LO}[\pi\pi]}\xspace}

\newcommand{\amuhadLBL}{\ensuremath{\amu^{\rm had,LBL}}\xspace}
\newcommand{\amuSM}{\ensuremath{\amu^{\rm SM}}\xspace}
\newcommand{\amuExp}{\ensuremath{\amu^{\rm exp}}\xspace}
\newcommand{\ol}{\ensuremath{\overline}}
\newcommand\cf{{cf.}\xspace}
\newcommand{\ea}{{\em et al.}\xspace}
\def\@citex[#1]#2{\if@filesw\immediate\write\@auxout{\string\citation{#2}}\fi
  \@tempcnta\z@\@tempcntb\m@ne\def\@citea{}\@cite{\@for\@citeb:=#2\do
    {\@ifundefined
       {b@\@citeb}{\@citeo\@tempcntb\m@ne\@citea
        \def\@citea{,\penalty\@m\ }{\bf ?}\@warning
       {Citation `\@citeb' on page \thepage \space undefined}}%
    {\setbox\z@\hbox{\global\@tempcntc0\csname b@\@citeb\endcsname\relax}%
     \ifnum\@tempcntc=\z@ \@citeo\@tempcntb\m@ne
       \@citea\def\@citea{,\penalty\@m}
       \hbox{\csname b@\@citeb\endcsname}%
     \else
      \advance\@tempcntb\@ne
      \ifnum\@tempcntb=\@tempcntc
      \else\advance\@tempcntb\m@ne\@citeo
      \@tempcnta\@tempcntc\@tempcntb\@tempcntc\fi\fi}}\@citeo}{#1}}

\def\@citeo{\ifnum\@tempcnta>\@tempcntb\else\@citea
  \def\@citea{,\penalty\@m}%
  \ifnum\@tempcnta=\@tempcntb\the\@tempcnta\else
   {\advance\@tempcnta\@ne\ifnum\@tempcnta=\@tempcntb \else
\def\@citea{--}\fi
    \advance\@tempcnta\m@ne\the\@tempcnta\@citea\the\@tempcntb}\fi\fi}
\newenvironment{myquote}
               {\list{}{\leftmargin0cm\indent}%
                \item\relax}
               {\endlist}
\newcommand\allFontSize{\footnotesize}
\newcommand\detailsSize{\allFontSize}
{\begin{myquote}\detailsSize}{\end{myquote}}

\begin{document}
 
\twocolumn[{%
  \begin{@twocolumnfalse}

    \begin{flushright}
      \normalsize
      \today
    \end{flushright}

    \vspace{-2cm}

    \title{\Large\boldmath A new evaluation of the hadronic vacuum polarisation contributions to the muon anomalous magnetic moment  and to \sansmath$\mathbf{\boldsymbol\alpha(m_Z^2)}$}

    \author{M.~Davier\inst{1} \and 
        A.~Hoecker\inst{2} \and 
        B.~Malaescu\inst{3} \and 
        Z.~Zhang\inst{1}}
 
    \institute{IJCLab, Universit\'e Paris-Saclay et CNRS/IN2P3, Orsay, France \and
          CERN, CH--1211, Geneva 23, Switzerland \and
          LPNHE, Sorbonne Universit\'e, Paris Diderot Sorbonne Paris Cit\'e, CNRS/IN2P3, Paris, France }
          
    \abstract{
      We reevaluate the hadronic vacuum polarisation contributions to the muon magnetic anomaly and to the running of the electromagnetic coupling constant at the $Z$-boson mass. We include newest $\ee \to {\rm hadrons}$ cross-section data together with a phenomenological fit of the threshold region in the evaluation of the dispersion integrals. The precision in the individual datasets cannot be fully exploited due to discrepancies that lead to additional systematic uncertainty in particular between BABAR and KLOE data in the dominant \pipi channel. For the muon $(g-2)/2$, we find for the lowest-order hadronic contribution  $(694.0 \pm 4.0)\cdot10^{-10}$. The full Standard Model prediction  differs by $3.3\sigma$ from the experimental value. The five-quark hadronic  contribution to \aZ is evaluated to be $(276.0\pm1.0)\cdot10^{-4}$. 
    }

    \maketitle
  \end{@twocolumnfalse}
}]

\section{~Introduction}
\label{sec:Introduction}

The Standard Model (SM) predictions of the anomalous magnetic moment of the muon, 
$\amu=(g_\mu-2)/2$, with $g_\mu$ the muon gyromagnetic factor, 
and of the running electromagnetic coupling constant, $\alpha(s)$,  an important  ingredient 
of electroweak theory, are 
limited in precision by    hadronic vacuum polarisation (HVP) contributions. 
The dominant hadronic terms can be calculated with the use of experimental 
cross-section data, involving \ee annihilation to hadrons, and perturbative QCD to evaluate  energy-squared dispersion integrals ranging from the $\piz\gamma$ 
threshold to infinity. The  kernels occurring in these integrals 
emphasise low photon virtualities, owing to the $1/s$ descent of the cross section, 
and, in case of \amu, to an additional $1/s$ suppression. About 73\% 
of the lowest order hadronic contribution to \amu and 58\% of the total uncertainty-squared
are given by the $\ppg$ final state,\footnote
{Throughout this paper, final state photon radiation is implied for all 
   hadronic final states.
} 
while this channel amounts to only 12\% of the hadronic contribution to $\alpha(s)$ 
at $s=\mZ^2$~\cite{dhmz2017}.

In this work, we reevaluate the lowest-order hadronic contribution, \amuhadLO, to 
the muon magnetic anomaly, and the hadronic contribution, \dahadZ, to the running 
\aZ at the $Z$-boson mass using newest $\ee \to {\rm hadrons}$ cross-section 
data and updated techniques. In particular, we perform a phenomenological fit to supplement less precise data in the low-energy domain up to 0.6$\;$GeV. We also reconsider the systematic uncertainty in the \pipi channel in view of discrepancies among the most precise datasets.

All the experimental contributions are evaluated using the software package
HVPTools~\cite{g209}. To these are added  narrow resonance contributions evaluated 
analytically, and continuum contributions computed using perturbative QCD. 

\section{Combination of experimental inputs}
\label{Sec:Combination}

\sloppy
The integration of data points belonging to different experiments with their own data densities requires a careful treatment to avoid biases and to properly account for correlated systematic uncertainties within the same experiment and between different experiments, as well as within and between different channels. Quadratic interpolation (splines) of adjacent data points is performed for each experiment, and a local combination in form of a weighted average of the interpolations is computed in bins of 1$\;$MeV, or in narrower bins for the $\omega$ and $\phi$ resonances. 

The  uncertainties on the combined dataset, the data integration and the phenomenological fit are computed using large numbers of pseudo-experiments. 
These are generated taking into account all measurement uncertainties and their correlations. While this treatment guarantees a proper propagation of uncertainties, the resulting precision of the combination still depends on the chosen test statistic: a poor choice (e.g., an arithmetic instead of a weighted average) would lead to poor precision, while an aggressive choice (e.g., exploiting the available correlation information globally over the full spectrum, thereby benefiting from constraints among different energy regimes\footnote{Systematic uncertainties are based on estimates which are impacted by imponderables regarding size and correlation among measurements, in particular  uncertainties due to theoretical  modelling. Systematic uncertainties are often evaluated in relatively wide mass ranges, the event topology may evolve between measurements performed at different centre-of-mass energies (affecting for example the acceptance and tracking efficiency) as does the background composition, systematic uncertainties due to  trigger and tracking may be correlated, etc. It is therefore important to treat systematic uncertainties and their correlations with care and avoid the use of long-range correlations to constrain measurements among different centre-of-mass energies. Ambiguities in systematic uncertainties and their correlations have been studied in other experimental areas and  different ``configurations"/``scenarios" of uncertainties were proposed~\cite{Aad:2014bia, Aaboud:2017dvo, Aaboud:2017wsi}. }) could lead to an optimistic precision claim with the risk of undercoverage with respect to the (unknown) truth. 
To avoid either case, we employ a test statistic that only relies on local measurement uncertainties and correlations to combine datasets in a given bin.\footnote{This information on the uncertainties and correlations is used on slightly wider ranges, of typically up to a couple of 100~MeV, when {\it averaging regions} are defined in order to account for the difference between the point-spacing and bin-sizes for the various experiments~\cite{g209}. In this procedure the systematic uncertainties are not constrained, but rather directly propagated from each input measurement to the {\it averaging regions} and then to the fine bins.}
As stated above, the uncertainty in each combined bin and the correlation among bins are evaluated using pseudo-experiments generated with the full correlation information. Correlations between channels are accounted for by propagating individually the common systematic uncertainties.\footnote{A number of 15 such uncertainties are accounted for in the current study. Typical examples are the luminosity uncertainties, if the data stem from the same experimental facility but measure different channels, and uncertainties related to radiative corrections.}

Where results from different datasets are locally inconsistent, the combined uncertainty is rescaled according to the local $\chi^2$ value and number of degrees of freedom following the PDG prescription~\cite{pdg}. Such inconsistencies are currently limiting the precision of the combination in the dominant $\pi^+\pi^-$ channel as well as in the $\Kp\Km$ channel  (see discussions below). In most exclusive channels the largest weight in the combination is provided by BABAR data. 

Closure tests with known distributions have been performed in the dominant \pipi channel to validate both the combination and integration procedures.

\section{~Input data}

Exclusive bare hadronic cross-section measurements for 32 channels are integrated up to 1.8$\;$GeV over the relevant dispersion kernels. This analysis uses all the available public data with recent additions~\cite{CMD-3:2017tgb,TheBaBar:2017vzo,Achasov:2017vaq,cleo2017,TheBABAR:2018vvb,Achasov:2018ujw,Lees:2018dnv,cmd3-7pi,Achasov:2019duv,Ivanov:2019crp}. References for data already included in our 2017 analysis are provided in the corresponding paper~\cite{dhmz2017} as well as earlier publications~\cite{dhmz2011,dehz2003}.

In the energy range 1.8--3.7$\;$GeV and above 5$\;$GeV four-loop perturbative QCD is used~\cite{baikov}. The contributions from the open charm pair production region between 3.7 and 5$\;$GeV are again computed using experimental data. For the narrow resonances $J/\psi$ and $\psi(2S)$ Breit-Wigner line shapes are integrated using their currently best known parameters~\cite{pdg}.

The following  discussion of individual channels  focuses on the HVP contribution to $\amu$ as it relies more strongly on the low-energy experimental data. We mainly explore the impact of the  data released since our last publications~\cite{dhmz2017,dhmz2011}. If not stated otherwise, all numerical results for \amu are quoted in units of $10^{-10}$.

\begin{figure*}[p]
\begin{center}
\includegraphics[width=130mm]{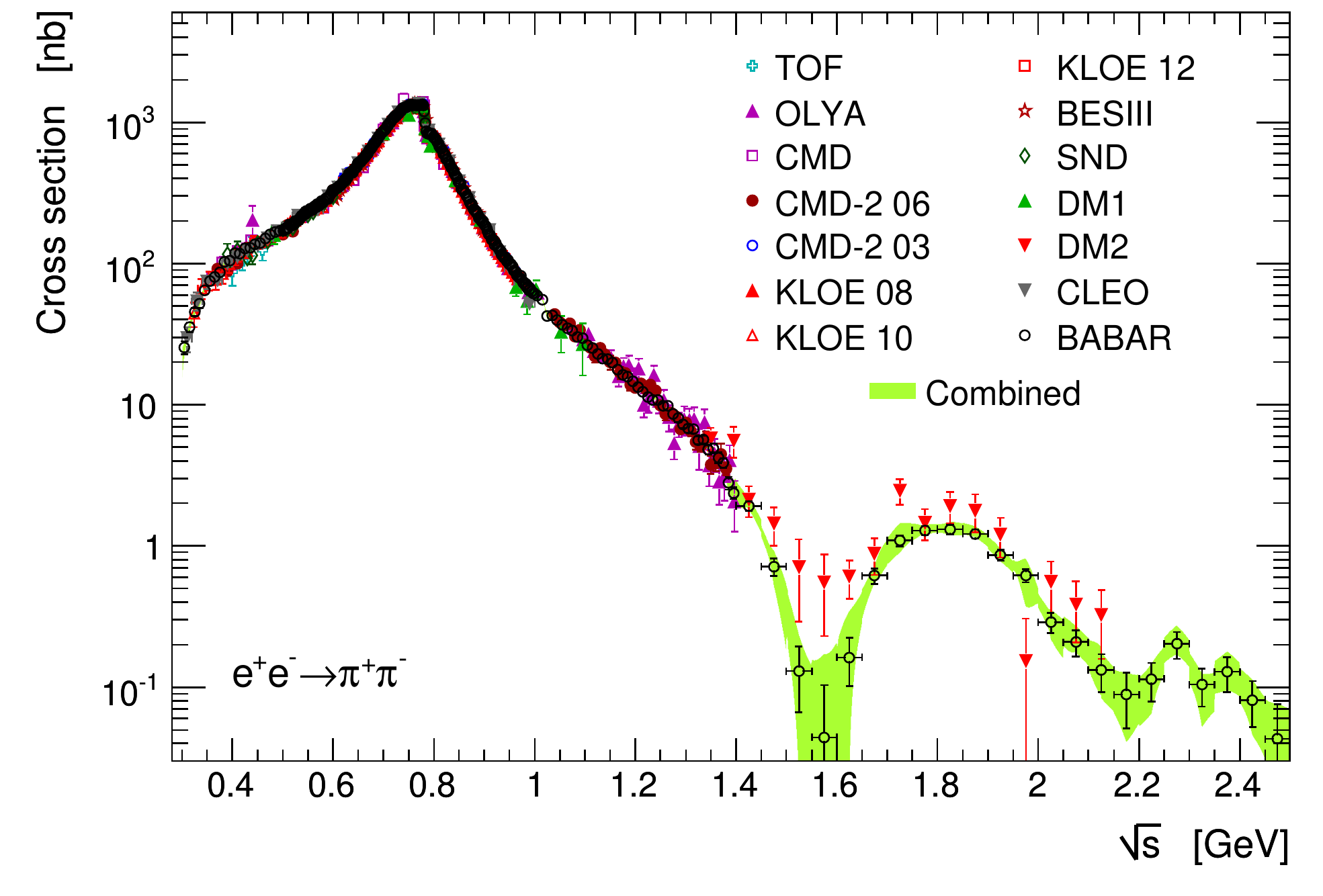}
\vspace{0.5cm}

\includegraphics[width=\figsize]{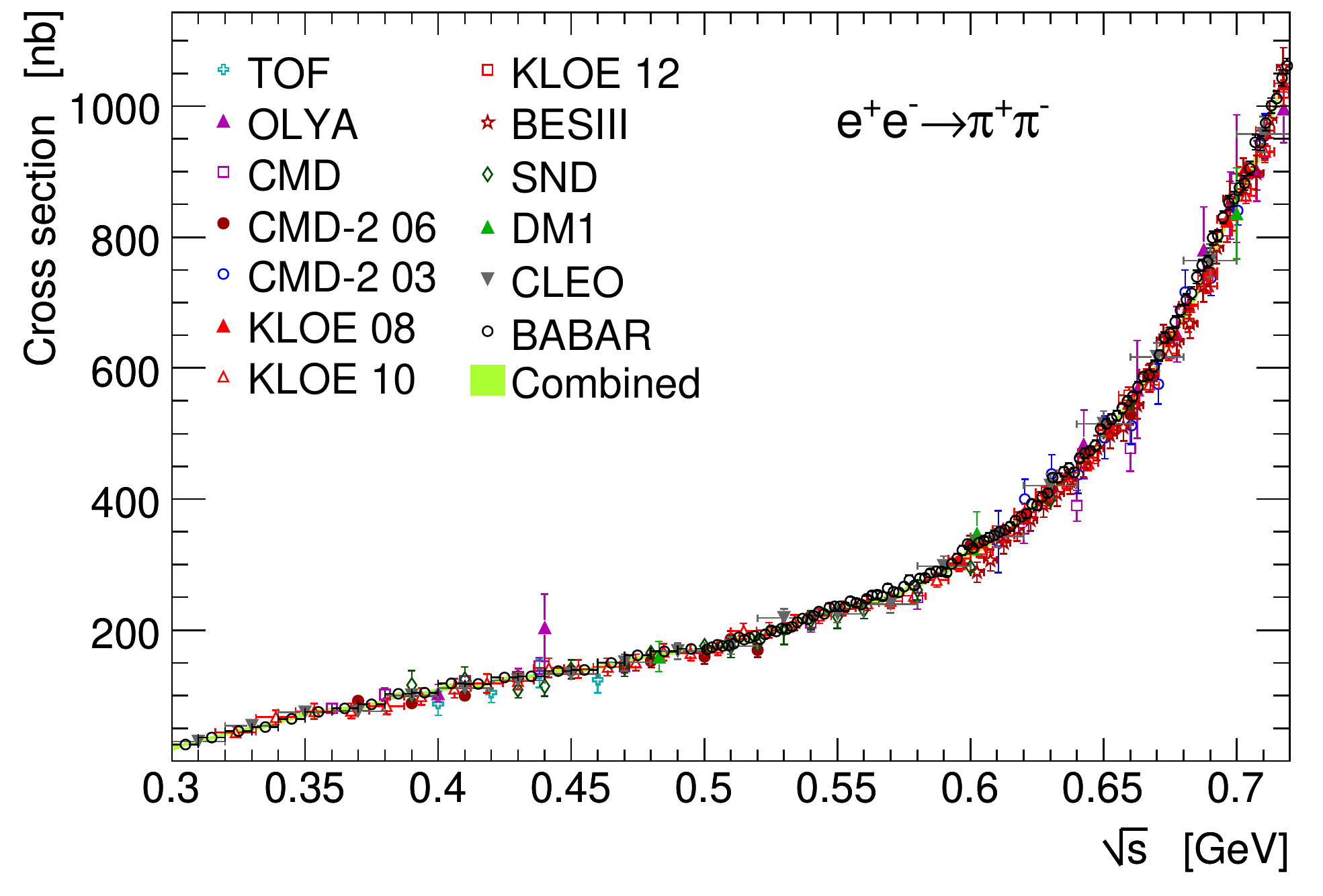}\hspace{\fighspace}
\includegraphics[width=\figsize]{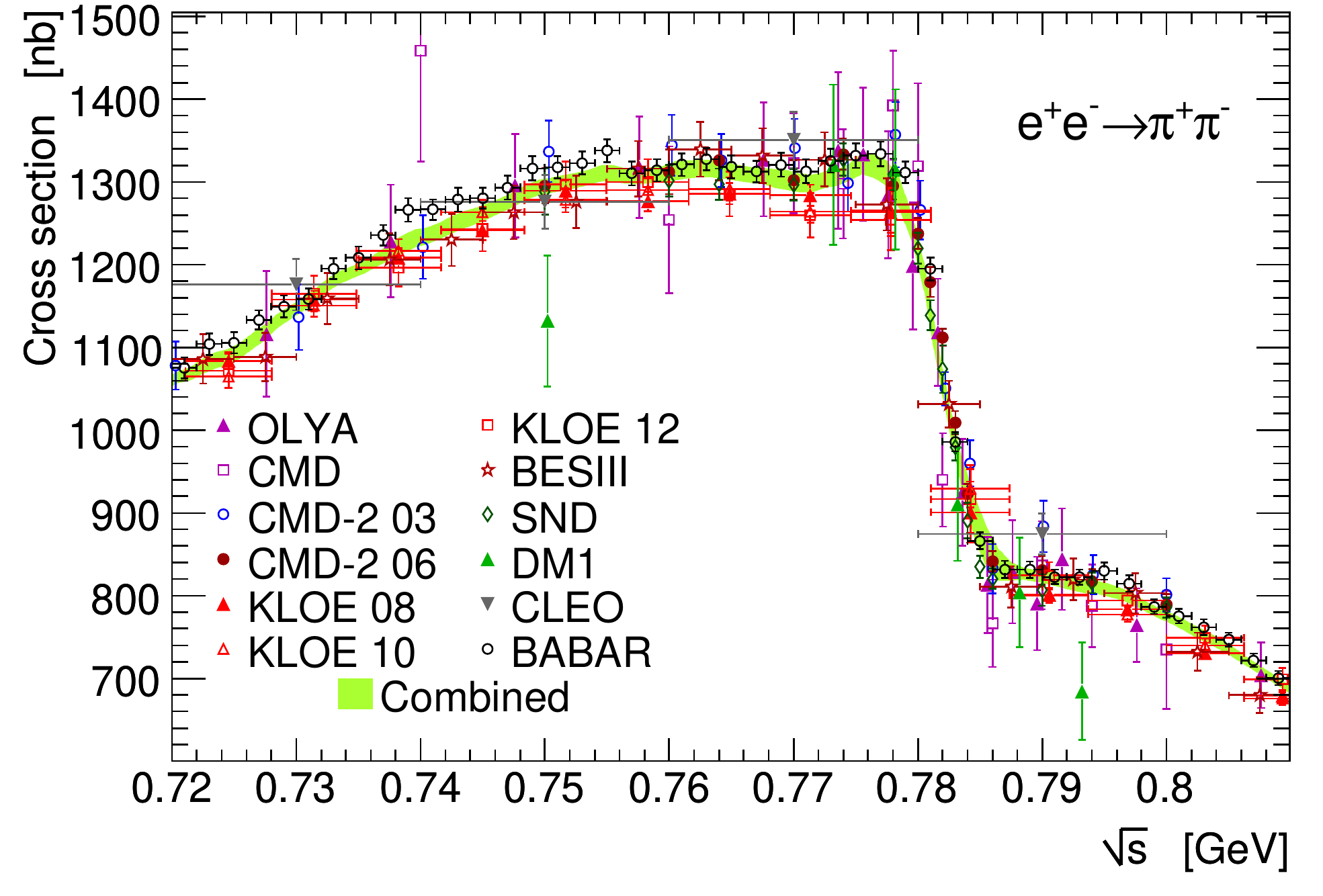}
\vspace{0.2cm}

\includegraphics[width=\figsize]{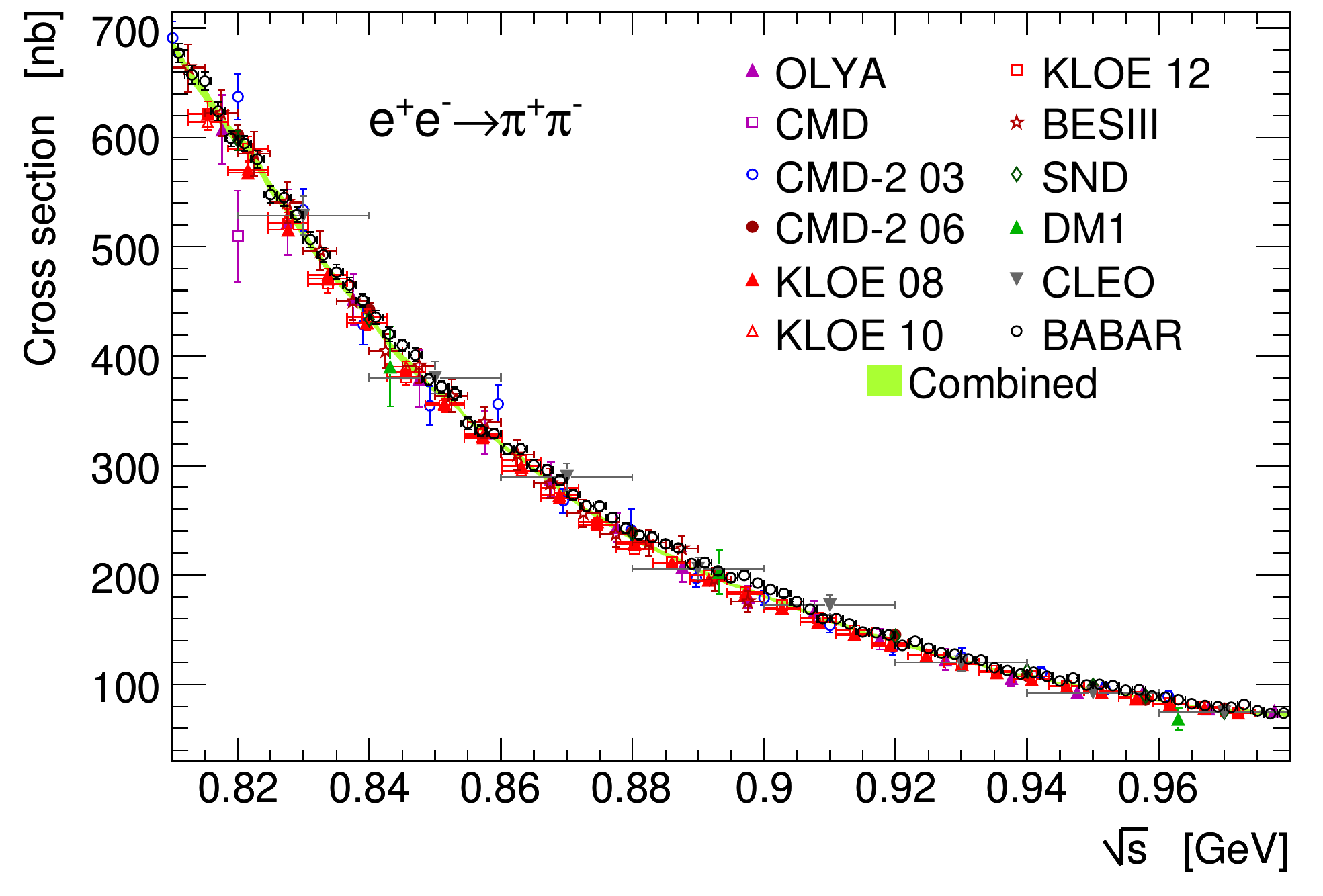}\hspace{\fighspace}
\includegraphics[width=\figsize]{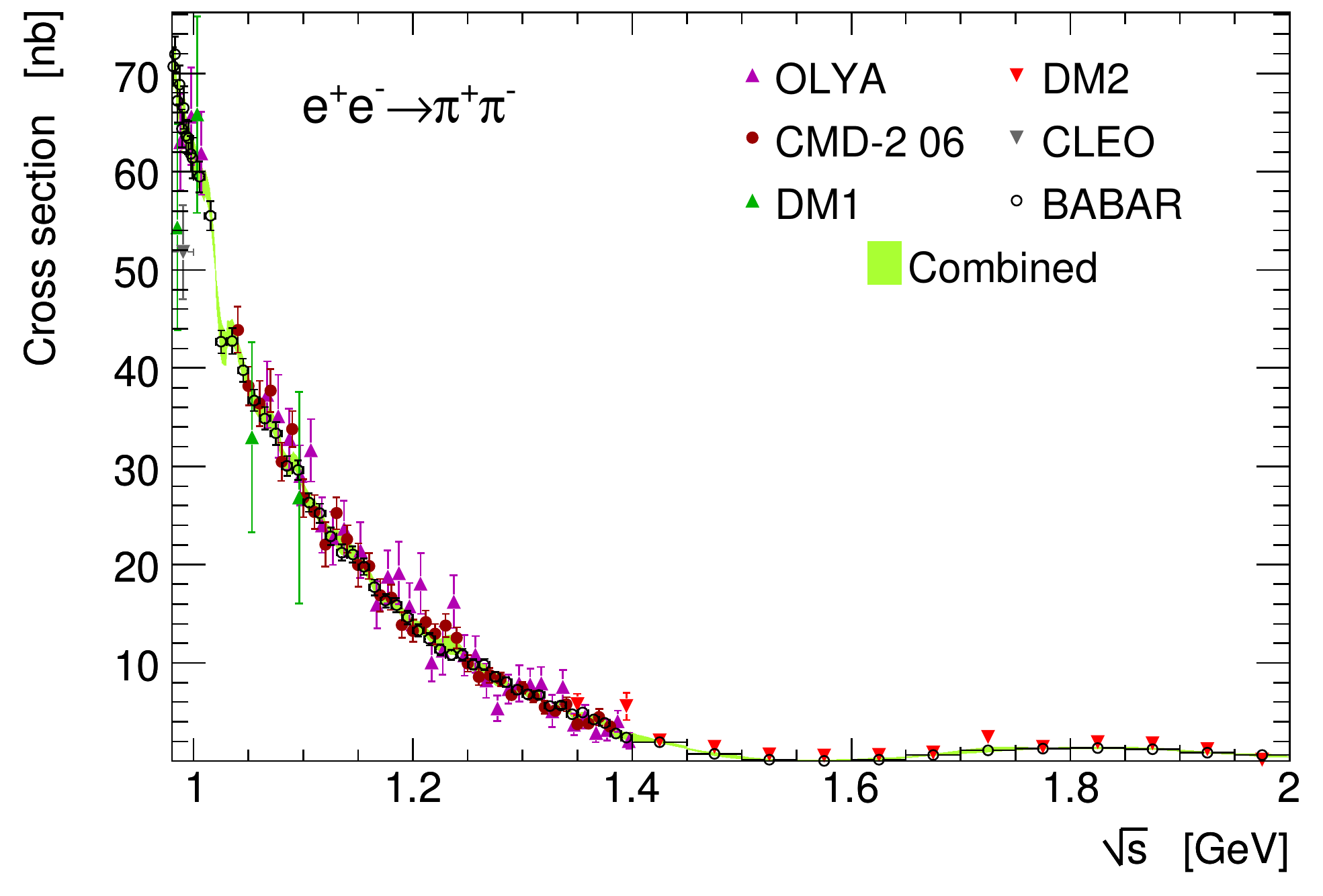}
\end{center}
\vspace{-0.2cm}
\caption[.]{ 
            Bare cross section of $\ee\to\pp$ versus centre-of-mass energy for different 
            energy ranges.  The error bars of the data points include statistical and systematic 
            uncertainties added in quadrature. The green band shows
            the HVPTools combination within its $1\sigma$ uncertainty. 
    
}
\label{fig:pipiall}
\end{figure*}
\begin{figure*}[p]
\begin{center}
\includegraphics[width=\figsize]{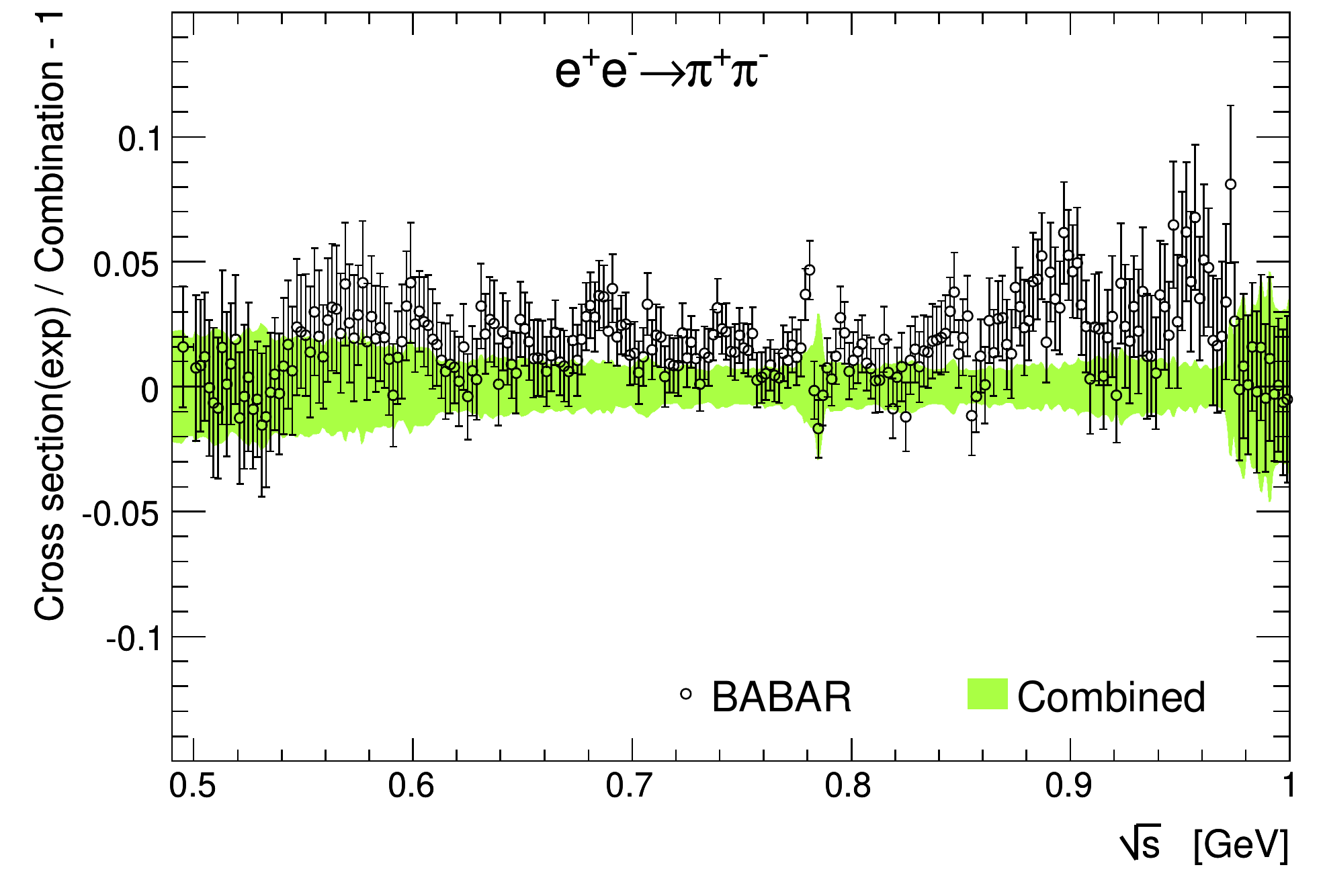}\hspace{\fighspace}
\includegraphics[width=\figsize]{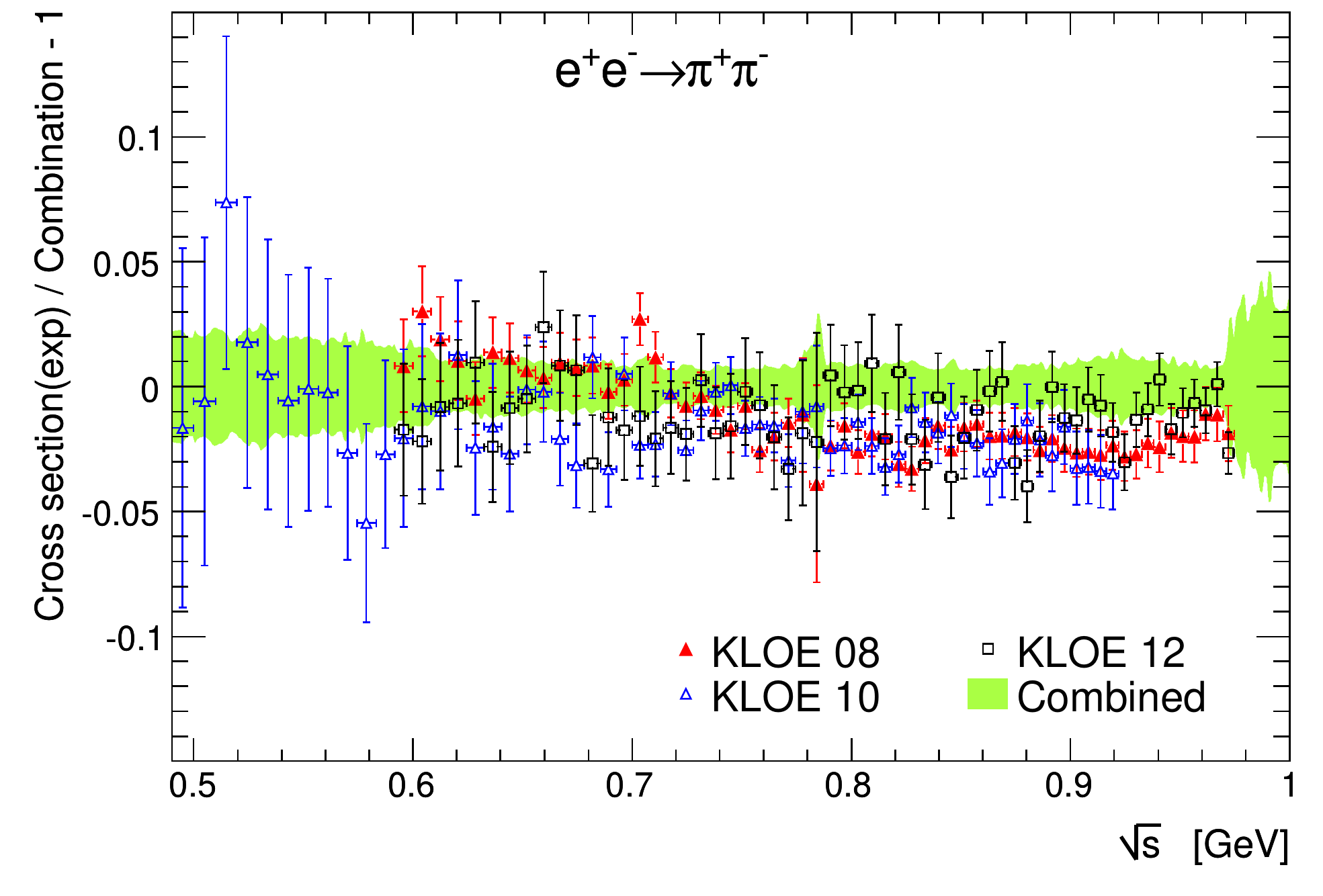}
\vspace{0.2cm}

\includegraphics[width=\figsize]{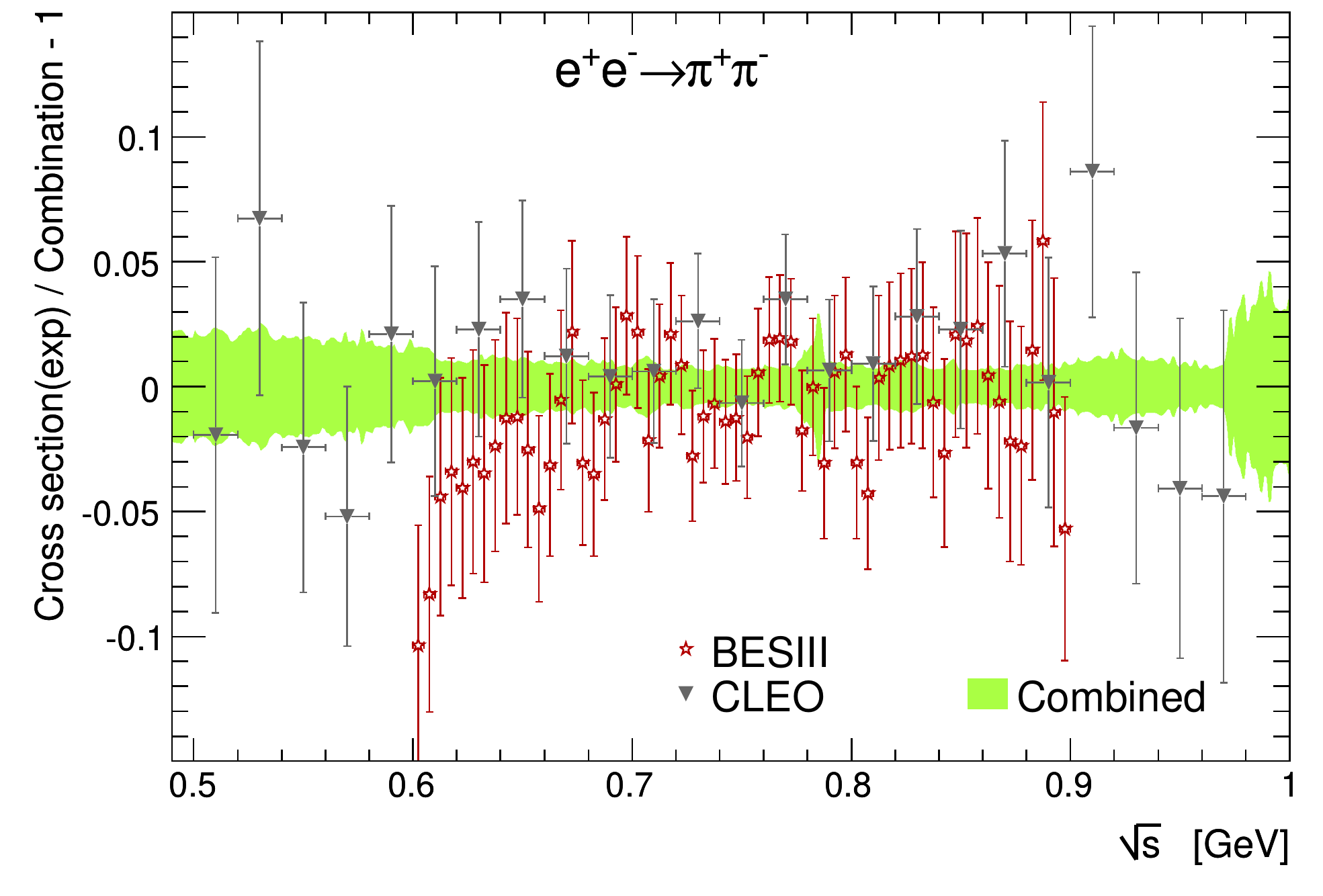}\hspace{\fighspace}
\includegraphics[width=\figsize]{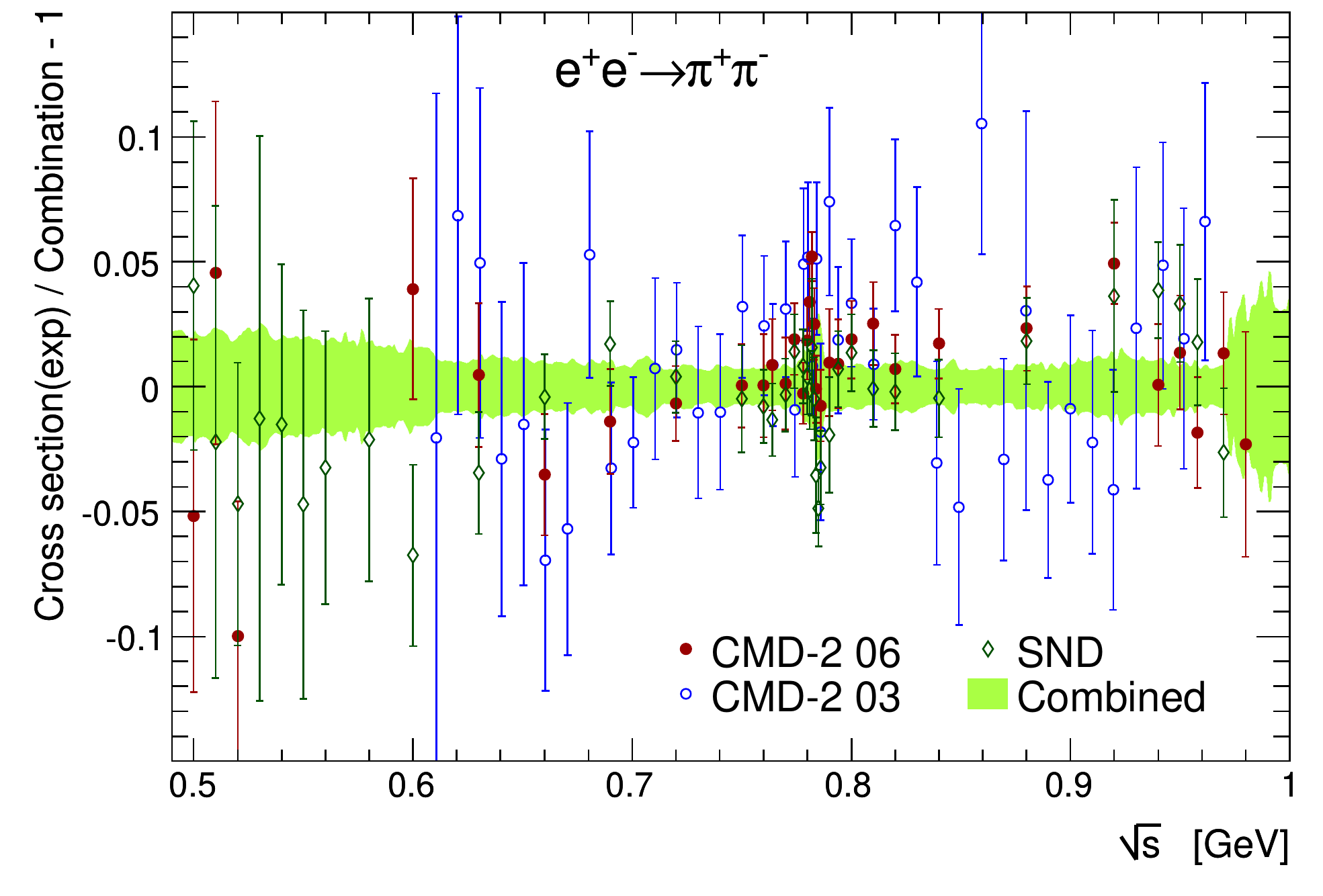}
\end{center}
\vspace{-0.2cm}
\caption[.]{ 
            Comparison between individual $\ee\to\pp$ cross-section measurements from 
            BABAR~\cite{babarpipi1,babarpipi2},  KLOE\,08~\cite{kloe08}, KLOE\,10~\cite{kloe10},
            KLOE\,12~\cite{kloe12}, BESIII~\cite{bes2015}, CLEO~\cite{cleo2017},
            CMD-2\,03~\cite{cmd203}, CMD-2\,06~\cite{cmd2new}, SND~\cite{snd2pi}, and 
            the HVPTools combination. The error bars include statistical and systematic 
            uncertainties added in quadrature. 
}
\label{fig:comppipi}
\vspace{1.5cm}
\includegraphics[width=\figsize]{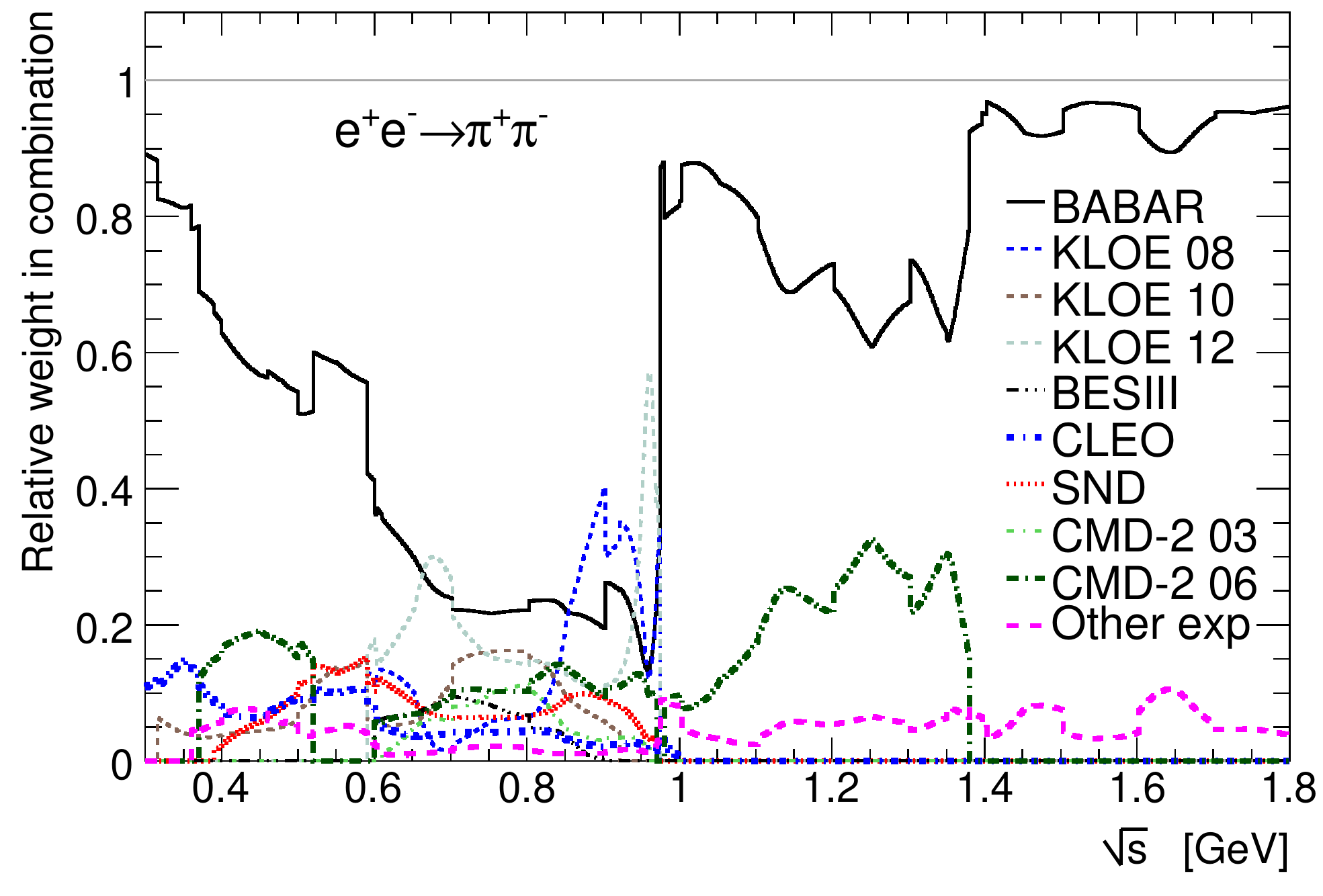}\hspace{\fighspace}
\includegraphics[width=\figsize]{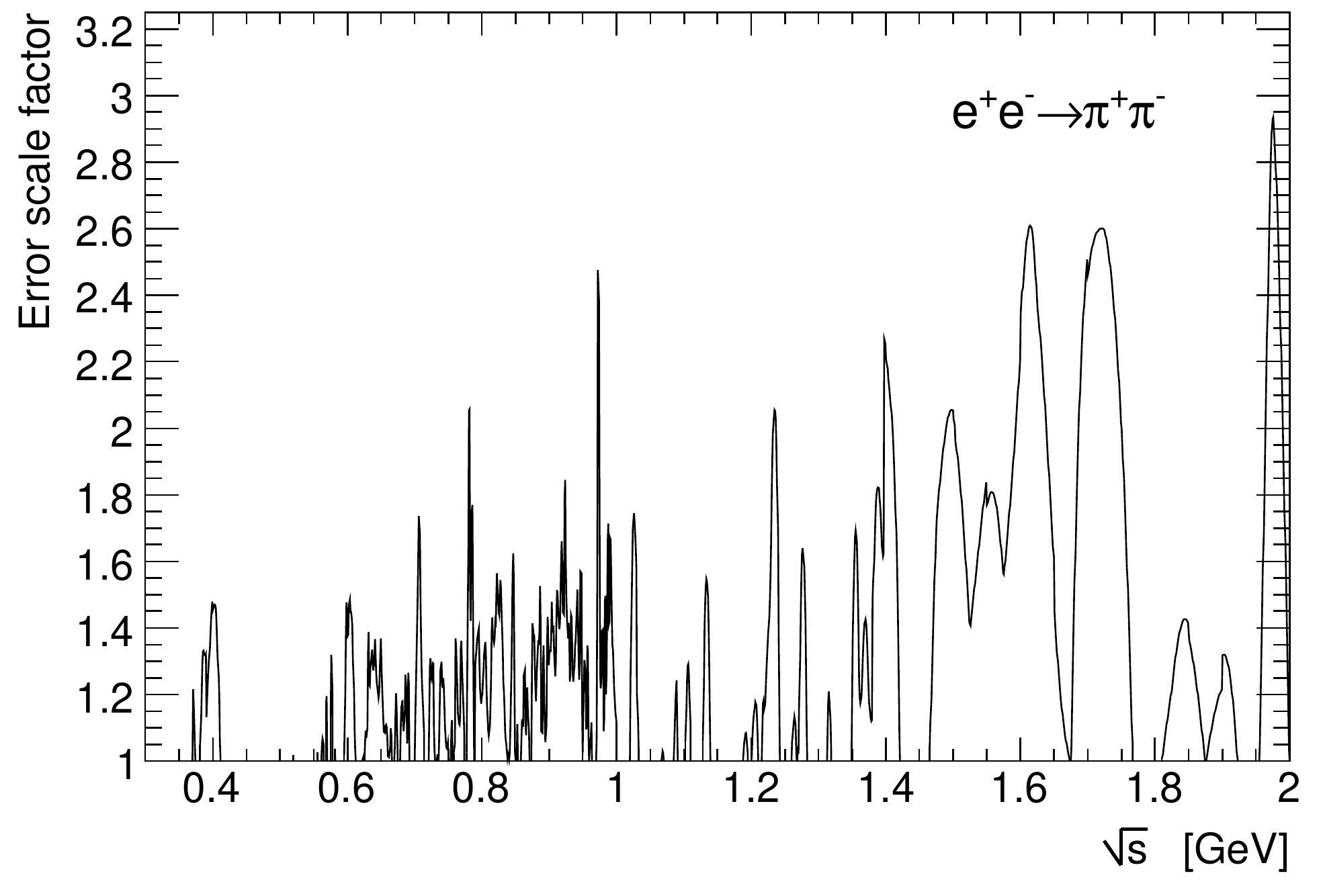}
\vspace{0.1cm}

\caption[.]{ 
            Left: relative local weight per experiment contributing to the $\ee\to\pp$
            cross-section combination versus centre-of-mass energy. 
            Right: local scale factor versus centre-of-mass energy applied 
            to the combined \pp cross-section uncertainty to account for inconsistency 
            in the individual measurements. }
\label{fig:weights}
\end{figure*}

\subsection{~The \sansmath$\mathbf{\boldsymbol{\pi^+\pi^-}}$ channel}
\label{sec:pipi}

Data from the BABAR~\cite{babarpipi1,babarpipi2} and KLOE~\cite{kloe08,kloe10,kloe12} experiments dominate  the measurement of the \pp channel. Their  sub-percent precision is not matched by the other experiments (CMD-2, SND, and BESIII\footnote{There is a small inconsistency between the bin-by-bin statistical uncertainties and the diagonal values of the statistical covariance matrix of the \pp data published by BESIII~\cite{bes2015}.}). New data in this channel stem from CLEO~\cite{cleo2017} using large angle initial state radiation (ISR) and taking into account up to one additional photon, following the BABAR method~\cite{babarpipi1}. Relatively large statistical uncertainties and a systematic uncertainty of 1.5\% are, however, insufficient to improve the precision of the combined $\pipi$ contribution. Recently, a combination of the three KLOE measurements was proposed~\cite{kloe17}. We do not use this combination as the KLOE measurements correspond to different ISR topologies and normalisation procedures.\footnote{Using the  KLOE combination~\cite{kloe17} we find for \amuhadLOpp  between the \pp threshold and $1.8\;$GeV a value of $506.6 \pm 2.4$, which is to be compared with $506.7 \pm 2.3$ as obtained from the HVPTools combination. For both calculations the usual local $\sqrt{\chiSdof}$ uncertainty rescaling method was applied. Without the rescaling the corresponding results are $506.6 \pm 2.0$ and $506.7 \pm 2.0$, respectively. The similarity of the results is maintained when using a phenomenological fit up to 0.6$\;$GeV (see later in text).} Eigenvector decomposition of the statistical and systematic covariance matrices of the three most recent series of KLOE measurements~\cite{kloe17} is used. Each eigenvector multiplied by the square-root of the corresponding eigenvalue is treated as an uncertainty source that is fully correlated between the  KLOE data points, while the individual sources are treated as independent among each other. Pseudo-experiments are generated in the usual way to propagate correlated uncertainties among the KLOE measurements.

Figure~\ref{fig:pipiall} shows the available $\ee\to\pp$ cross-section measurements in various panels zooming into different energy ranges. The green band indicates the HVPTools combination within its $1\sigma$ uncertainty. Comparisons between the combination and the most precise individual measurements are plotted in Fig.~\ref{fig:comppipi}. Figure~\ref{fig:weights} (left) shows the local combination weight versus $\sqrt{s}$ per experiment. The BABAR and KLOE measurements dominate over the entire energy range. Owing to the sharp radiator function, the KLOE event yield  increases towards the $\phi(1020)$ mass leading to a better precision than BABAR in the 0.8$-$1.0 GeV region. The group of experiments labelled ``Other exp'' in the left panel of Fig.~\ref{fig:weights} corresponds to older data with incomplete radiative corrections. Their weights are small throughout the entire energy domain. The right hand panel of Fig.~\ref{fig:weights} shows the scale factor versus centre-of-mass energy that is locally applied to the combined  \pp cross-section uncertainty to account for inconsistencies among the individual measurements. Significant inconsistencies are found between the most precise BABAR and KLOE datasets. 

The computation of the dispersion integral over the full \pp spectrum requires to extend the available data to the region between threshold and $0.3\:\gev$, for which we use a fit as described below.

\subsubsection*{Phenomenological fit}

\begin{figure*}[t]
\begin{center}
\includegraphics[width=\figsize]{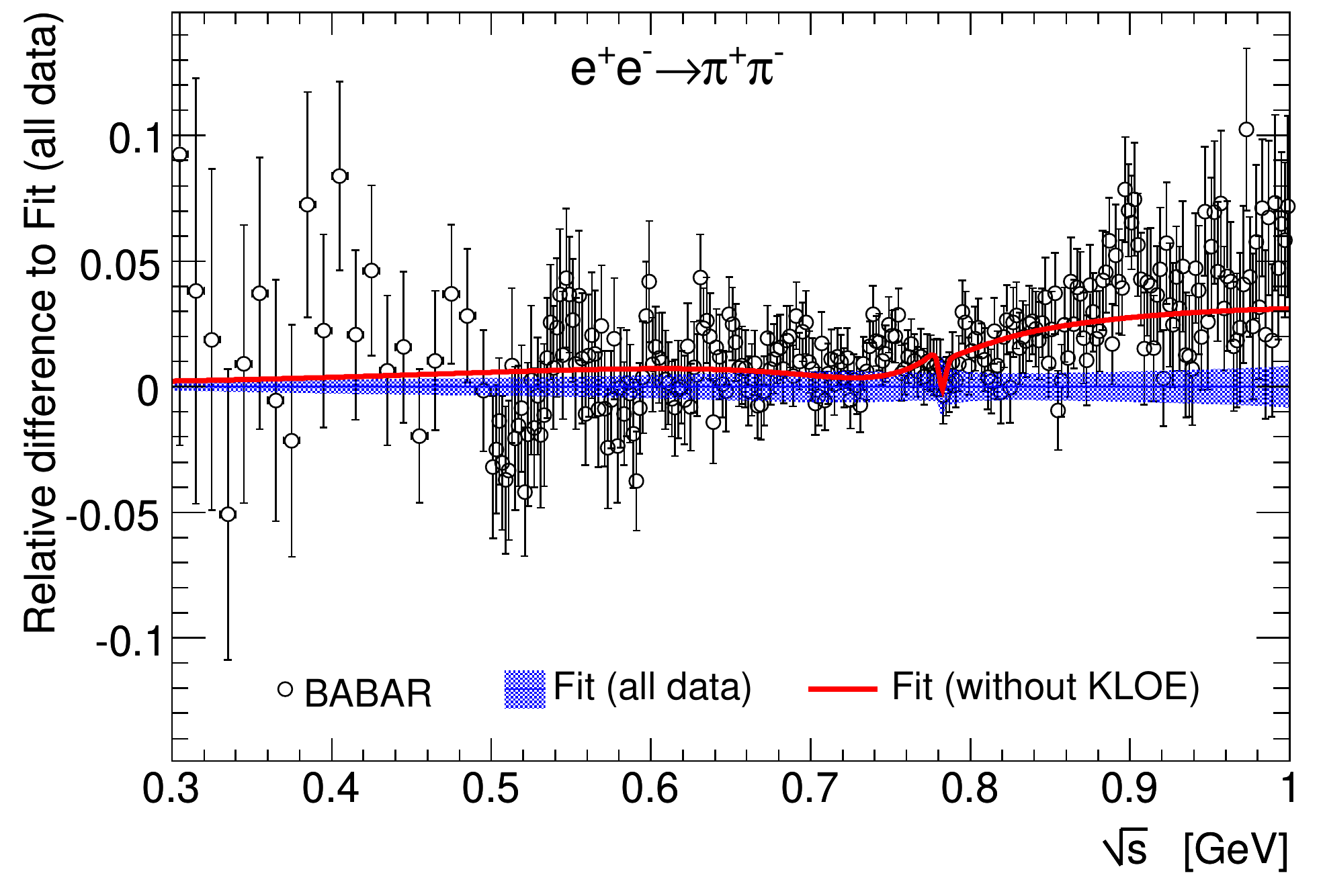}\hspace{\fighspace}
\includegraphics[width=\figsize]{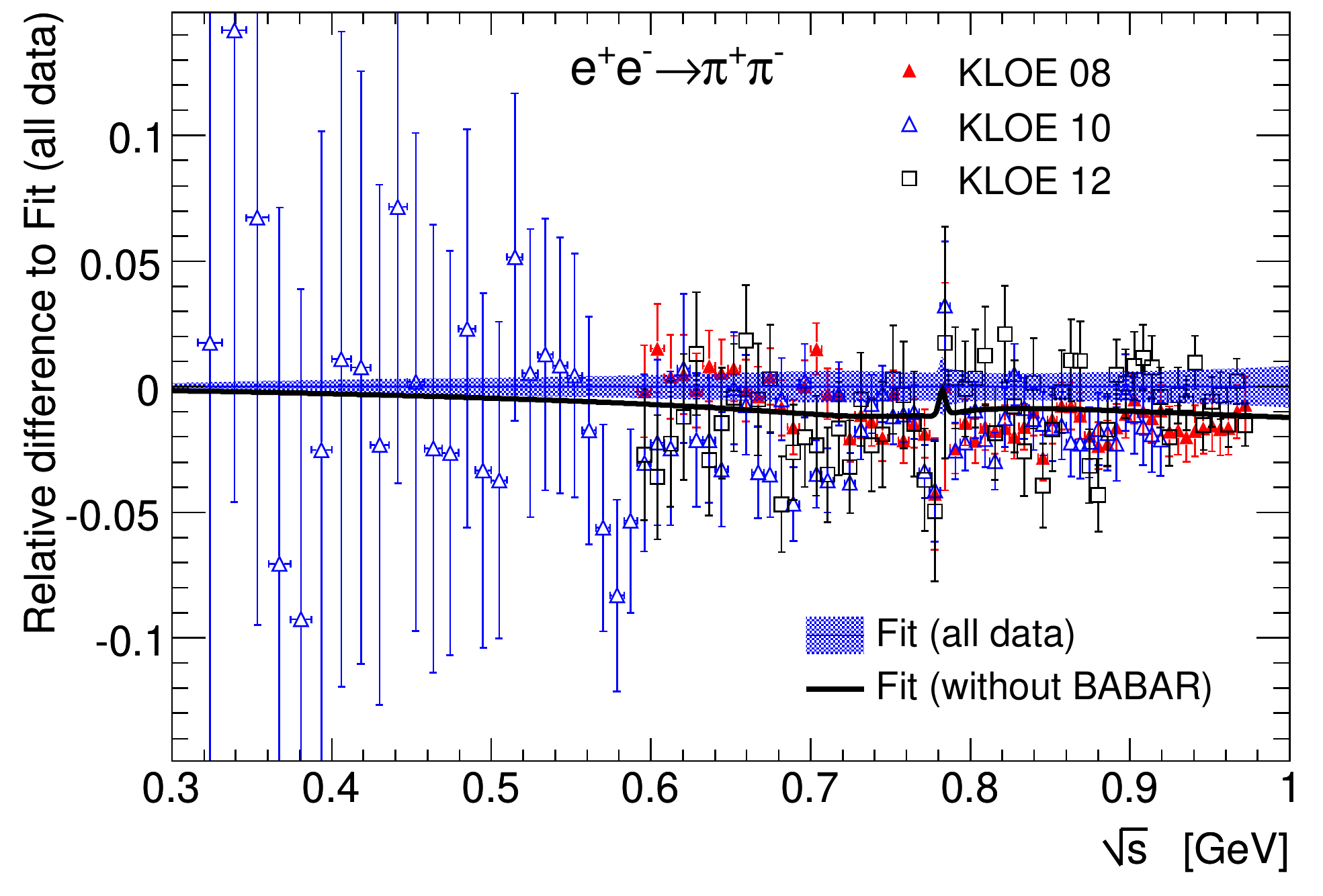}
\vspace{0.2cm}

\includegraphics[width=\figsize]{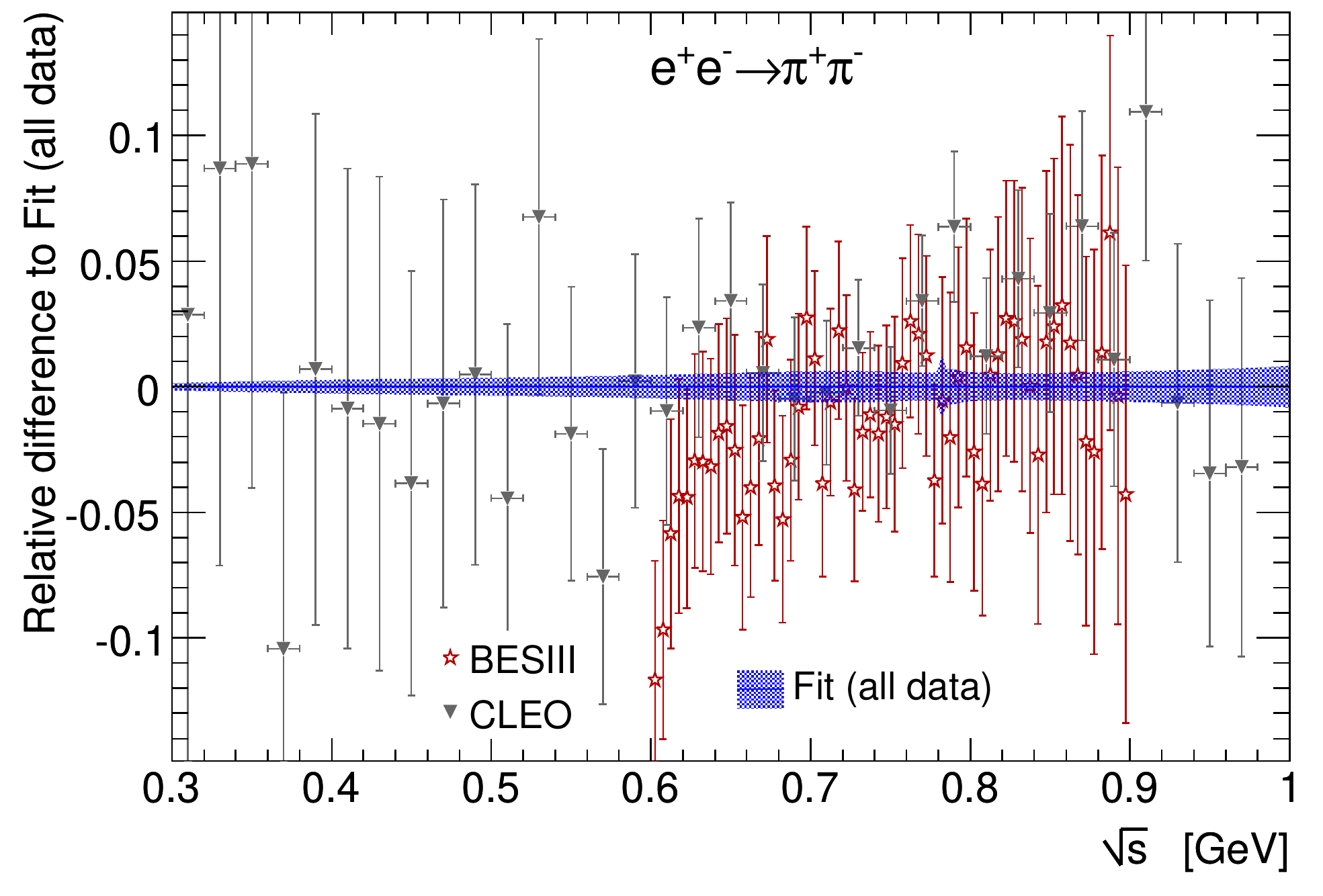}\hspace{\fighspace}
\includegraphics[width=\figsize]{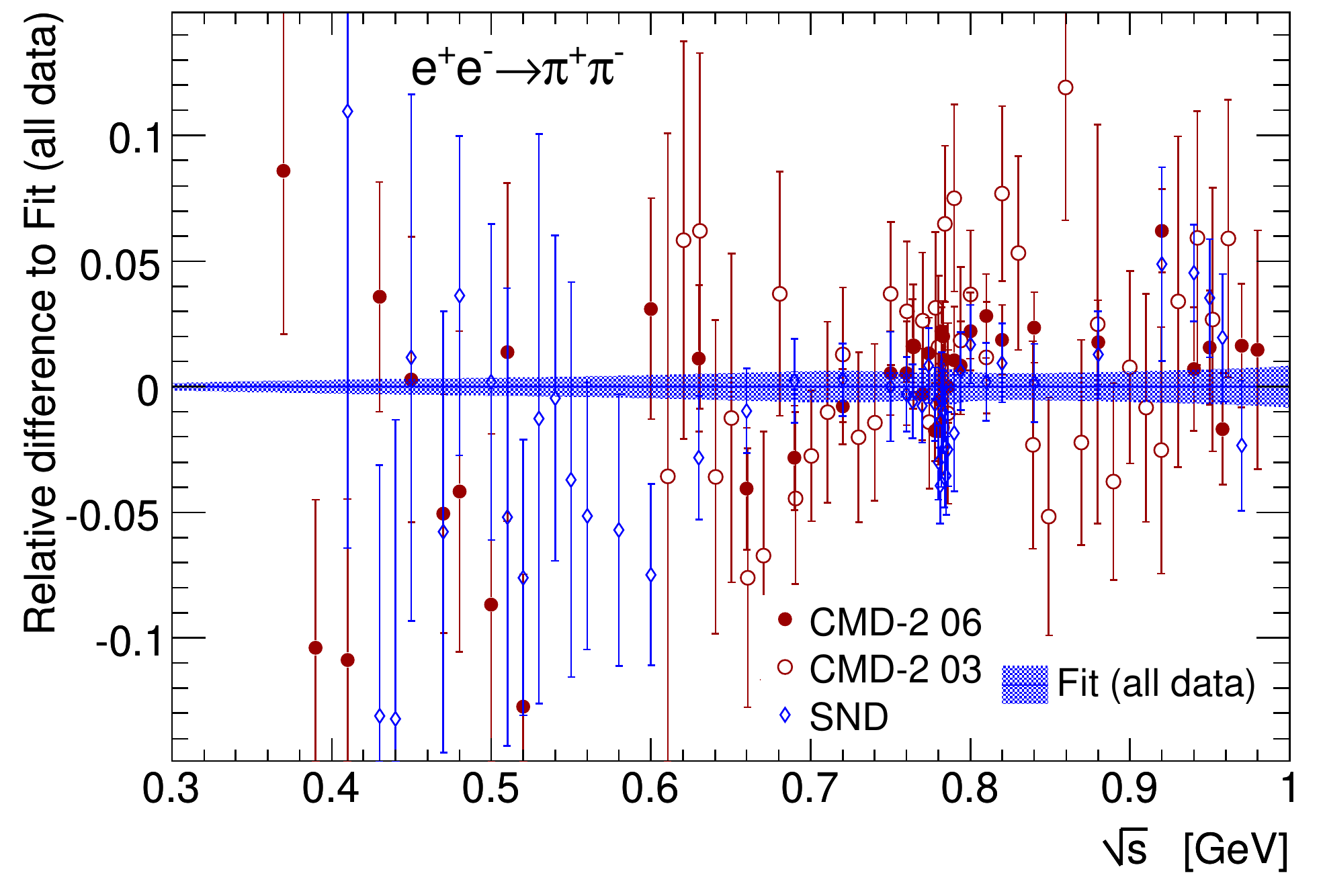}
\end{center}
\vspace{-0.2cm}
\caption[.]{ 
            Same as Fig.~\ref{fig:comppipi} except that the comparison is made with respect to the fit instead of the combination.
            The black and red curves show the results of two alternative fits where the data from KLOE and BABAR, respectively, were excluded.
}
\label{fig:data-fit}
\end{figure*}
The bare $e^+e^-\to\pp$ annihilation cross section is related to the pion form factor $F^0_\pi(s)$ (excluding vacuum polarisation) by
\begin{eqnarray}
    &&\sigma^{(0)}(e^+e^-\to \pi^+\pi^-)= \nonumber\\
      &&\hspace{1.5cm}\frac{\pi\alpha^2}{3s}\beta^3_0(s)\cdot|F^0_\pi(s)|^2\cdot\textrm{FSR}(s)\,,
\end{eqnarray}
where $\alpha$ is the electromagnetic coupling constant,  $\beta_0(s)=\sqrt{1-4m_\pi^2/s}$ is a threshold kinematic factor and FSR$(s)$ is the final state radiation contribution.

The pion form factor is an analytic function of $s$ in the complex plane, except on the real axis above $4m^2_\pi$. It can be parameterised as a product of two functions~\cite{Hanhart:2016pcd}
\begin{equation}
    F^0_\pi=G(s)\cdot J(s)\,,
    \label{Eq:FeRJ}
\end{equation}
where
\begin{eqnarray}
G(s)&=&1+\alpha_V s + \frac{\kappa s}{m^2_\omega -s -im_\omega\Gamma_\omega}\label{eq:rs}\,,
\end{eqnarray}
and, exploiting the unitarity constraint which identifies ${\rm arg}(F_\pi^0)$ with the P-wave \pp phase shift $\delta_1(s)$, 
\begin{eqnarray}
J(s)&=&e^{1-\delta_1(s_0)/\pi}\cdot\left(1-\frac{s}{s_0}\right)^{\!\!\left[1-\frac{\delta_1(s_0)}{\pi}\right]\frac{s_0}{s}}\!\!\left(1-\frac{s}{s_0}\right)^{\!\!-1}\nonumber\\
&& \cdot\; {\rm exp}\left({\frac{s}{\pi}\int^{s_0}_{4m^2_\pi}dt\frac{\delta_1(t)}{t(t-s)}}\right).\label{eq:js}
\end{eqnarray}
The last term in Eq.~(\ref{eq:rs}) accounts for $\rho-\omega$ mixing. The function $J(s)$ is taken from Refs.~\cite{DeTroconiz:2001rip,deTroconiz:2004yzs}. Owing to $\rho$ dominance, the phase shift $\delta_1(s)$ can be parameterised by~\cite{GarciaMartin:2011cn}
\begin{equation}
    \cot\delta_1(s)=\frac{\sqrt{s}}{2k^3(s)}\left(m^2_\rho-s\right)\left(\frac{2m^3_\pi}{m^2_\rho\sqrt{s}}+B_0+B_1\omega(s)\right)\label{eq:phase}
\end{equation}
with
\begin{eqnarray}
&& k(s)=\frac{\sqrt{s-4m^2_\pi}}{2}\nonumber\,,~~~ \omega(s)=\frac{\sqrt{s}-\sqrt{s_0-s}}{\sqrt{s}+\sqrt{s_0-s}}\,.
\end{eqnarray}
The six free parameters $\alpha_V$, $\kappa$, $m_\omega$, $m_\rho$, $B_0$ and $B_1$ are determined by the fit to the \pp data restricted to the region up to 1$\;$GeV to stay below the threshold of significant inelastic channels. The width of the $\omega$ resonance is fixed to its PDG value of 8.49$\;$MeV~\cite{pdg}, and $\sqrt{s_0}=1.05\:$GeV. The results of the fit are given in Table~\ref{tab:fit}. To derive an estimate for the model uncertainty, we independently vary $\sqrt{s_0}$ to 1.3$\;$GeV and remove the linear term $B_1\omega(s)$ from Eq.~(\ref{eq:phase}) since the resulting value of $B_1$ from the nominal fit is consistent with zero. 

The fit is performed using as test statistic a diagonal \chiS function that accounts for the statistical and systematic uncertainties of the experimental measurements.\footnote{For the fit we use the original data provided by each experiment instead of the HVPTools combination.} The same uncertainty rescaling in case of local discrepancies among datasets as for the HVPTools based combination is applied. Correlations are ignored in the test statistic, but accounted for in the uncertainty propagation through a series of pseudo-experiments for each of which the full fit procedure is repeated. This is a conservative procedure, as exploiting correlations in the test statistic would improve the precision of the fit.
Currently, the most precise measurements are dominated by systematic uncertainties, whose size and mass dependence as well as correlation among each other and among data points rely on estimates with somewhat limited precision, as discussed in section~\ref{Sec:Combination}.
Since there are also clear indications of a significant underestimate of the size of uncertainties in the discrepant dataset(s), we prefer not to exploit this information in the constrained fit.
Pseudo-experiments are also used to assess the goodness-of-fit on the data, which yields a p-value of 0.27.\footnote{The p-values for each individual dataset read 0.042 (BABAR), 0.097 (KLOE), 0.449 (CMD), 0.675 (TOF), 0.718 (DM1), 0.756 (CMD-2), 0.796 (SND), and 0.984 (CLEO). The p-values for both OLYA and BESIII are close to 1.}
We have checked the reliability of this procedure by generating a set of pseudo-experiments and evaluating the p-value for each of them.
The expected distribution of p-values reconstructed this way is indeed uniform between $0$ and $1$.

A graphical comparison of the fit result with the data is shown in Fig.~\ref{fig:data-fit}.
\begin{table*}[tbh]
\setlength{\tabcolsep}{0.0pc}
\begin{tabularx}{\textwidth}{@{\extracolsep{\fill}}lrrrrrr} 
\hline\noalign{\smallskip}
 & $\alpha_V$ & $\kappa [10^{-4}]$ & $B_0$ & $B_1$ & $m_\rho$ [MeV] & $m_\omega$ [MeV] \\
\noalign{\smallskip}\hline\noalign{\smallskip}
$\alpha_V$ & $0.133\pm 0.020$ & 0.52 & $-0.45$ & $-0.97$ & 0.90 & $-0.25$ \\
$\kappa [10^{-4}]$ & & $21.6\pm 0.5$ & $-0.33$ & $-0.57$ & 0.64 & $-0.08$ \\
$B_0$ & & & $1.040\pm 0.003$ & 0.40 & $-0.40$ & 0.29 \\
$B_1$ & & & & $-0.13\pm 0.11$ & $-0.96$ & 0.20 \\
$m_\rho$ [MeV] & & & & & $774.5\pm 0.8$ & $-0.17$ \\
$m_\omega$ [MeV] & & & & & & $782.0\pm 0.1$ \\
\noalign{\smallskip}\hline
\end{tabularx}
    \caption{Results of the fit to all $\pi^+\pi^-$ data. The diagonal elements give the fitted parameter values and their uncertainties, while the off-diagonal elements give the correlation coefficients.}
    \label{tab:fit}
\end{table*}
In the energy range between 0.3 and 0.6 GeV, the result of the fit yields for \amuhadLOpp a contribution of $109.8 \pm 0.4\pm 0.4$, where the first error is experimental and the second the model uncertainty. The latter is obtained by adding linearly the absolute values of following two variations: the $\sqrt{s_0}$ variation of $-0.13\pm 0.10$ and the difference of without and with the $B_1\omega(s)$ term of $0.24\pm 0.14$, where the uncertainty of each variation accounts for the correlation between the integral results. The corresponding result based on data integration is $109.6 \pm 1.0$. Taking into account the correlation of $72\%$ between the  experimental uncertainties, the difference between the two evaluations amounts to $0.2 \pm 0.9$. Similarly, for \dahadZ the difference is $0.020 \pm 0.028$. The fit therefore gives compatible but more precise results than the direct data integration.

Other studies using constraints from unitarity and analyticity with the aim to improve the precision of the \pp HVP contribution to the muon $g-2$ exist in the literature.\footnote{In Ref.~\cite{Hoferichter:2019gzf} an analyticity-based phenomenological fit has been used for the $\pi^+\pi^-\pi^0$ channel.}
The treatment followed in Ref.~\cite{Colangelo:2018mtw} is similar to ours with, however, a more elaborate theoretical analysis.
Differences are also present in the treatment of experimental data, Ref.~\cite{Colangelo:2018mtw} using a \chiS computed globally, including correlations across all the experimental data points and bins in the full mass range of interest. However, in order to avoid too low p-values, some bins of the KLOE measurements were removed in that study and energy rescaling parameters were introduced to fit each measured energy spectrum.
A different approach is followed in Ref.~\cite{Ananthanarayan:2018nyx,Ananthanarayan:2019zic} where the low-mass contribution was obtained from input data at a fixed mass, followed by a simple average used to combine inputs in the \sqrtS range between $0.65$ and $0.71$~GeV (and then to combine values from different experiments).
Instead of a direct evaluation of the correlations from the published information, they were assumed to be the same between all the combination inputs and an attempt was made to evaluate them based on the resulting \chiS value.
It is possible to directly compare the results for the mass region between threshold and 0.63 GeV. In the present analysis a value of $133.2 \pm 0.5 \pm 0.4$ is found, which agrees with the other results, $132.8 \pm 0.4 \pm 1.0$~\cite{Colangelo:2018mtw} and $132.9 \pm 0.8$~\cite{Ananthanarayan:2018nyx,Ananthanarayan:2019zic}.

It is also interesting to compare the results given in Table~\ref{tab:fit} with other analyses. The value obtained for $\kappa$ corresponds to a branching fraction of $\omega$ into $\pi^+\pi^-$ of $(2.09\pm0.09)\cdot10^{-2}$, in agreement with the result extracted from the fit of Ref~\cite{Colangelo:2018mtw}, $(1.95\pm0.08)\cdot10^{-2}$. Both values disagree with the PDG average~\cite{pdg}, $(1.51\pm0.12)\cdot10^{-2}$, dominated by the result of Ref.~\cite{Hanhart:2016pcd} which uses fits to essentially the same data. The fitted $\omega$ mass is found to be lower than the PDG average~\cite{pdg} obtained from $3\pi$ decays by $(0.65\pm0.12\pm0.12_{\rm PDG})$~MeV, in agreement with previous fits of the $\rho-\omega$ interference in the $2\pi$ spectrum (see for instance Refs.~\cite{babarpipi2,Colangelo:2018mtw}).

\subsubsection*{The \sansmath\pip\pim contribution}

\begin{figure}[t]
\vspace{0.4cm}
\begin{center}
\includegraphics[width=\figsize]{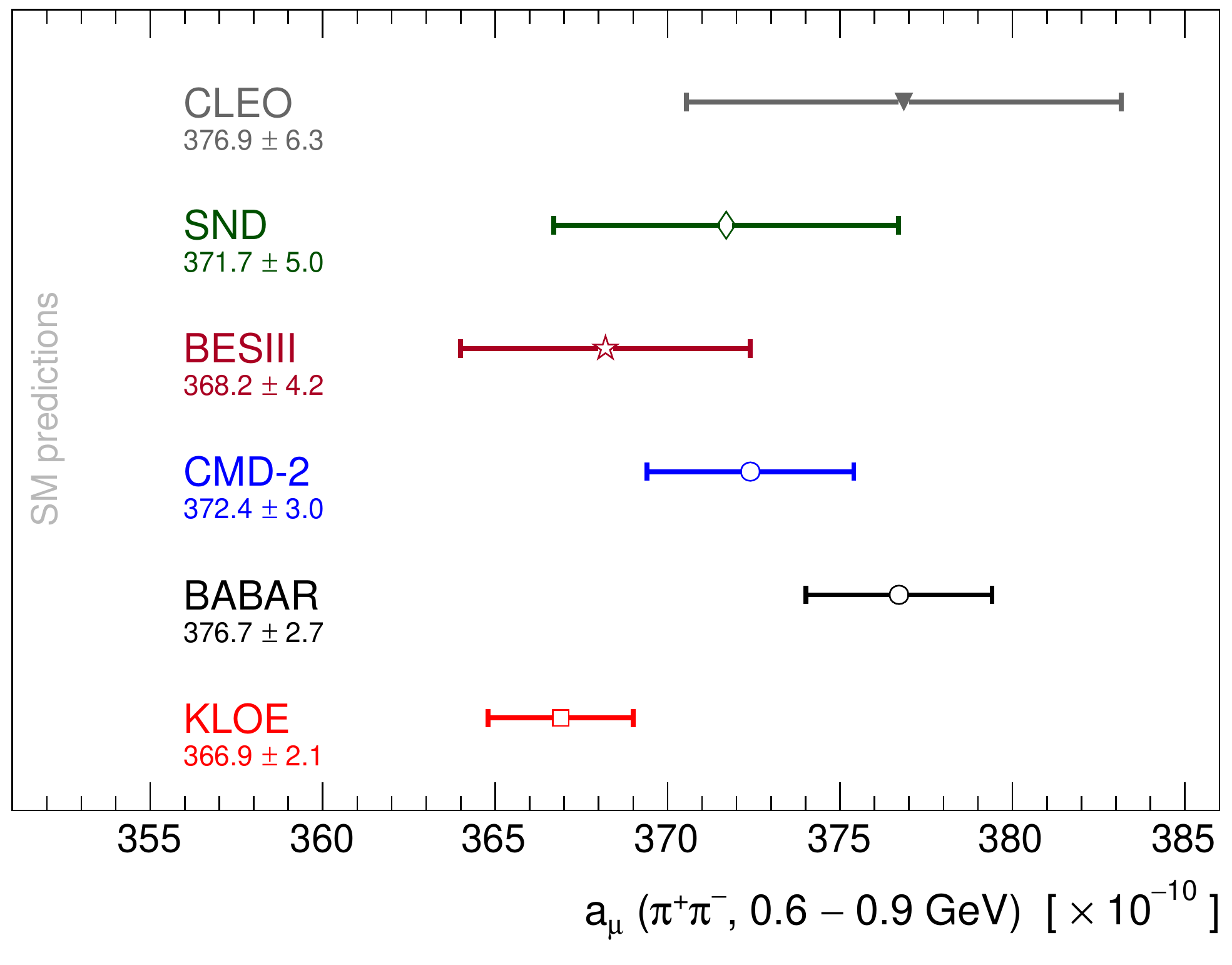}
\end{center}
  \vspace{-0.15cm}
  \caption{Comparison of results for \amuhadLOpp, evaluated between 0.6$\;$GeV and 0.9$\;$GeV for the various experiments. In case of CMD-2 all available measurements have been combined using HVPTools. For KLOE the result from the public combination~\cite{kloe17} is displayed.}
  \label{fig:amu2pi}
\end{figure}
\begin{figure}[t]
\vspace{0.4cm}
\begin{center}
\includegraphics[width=\figsize]{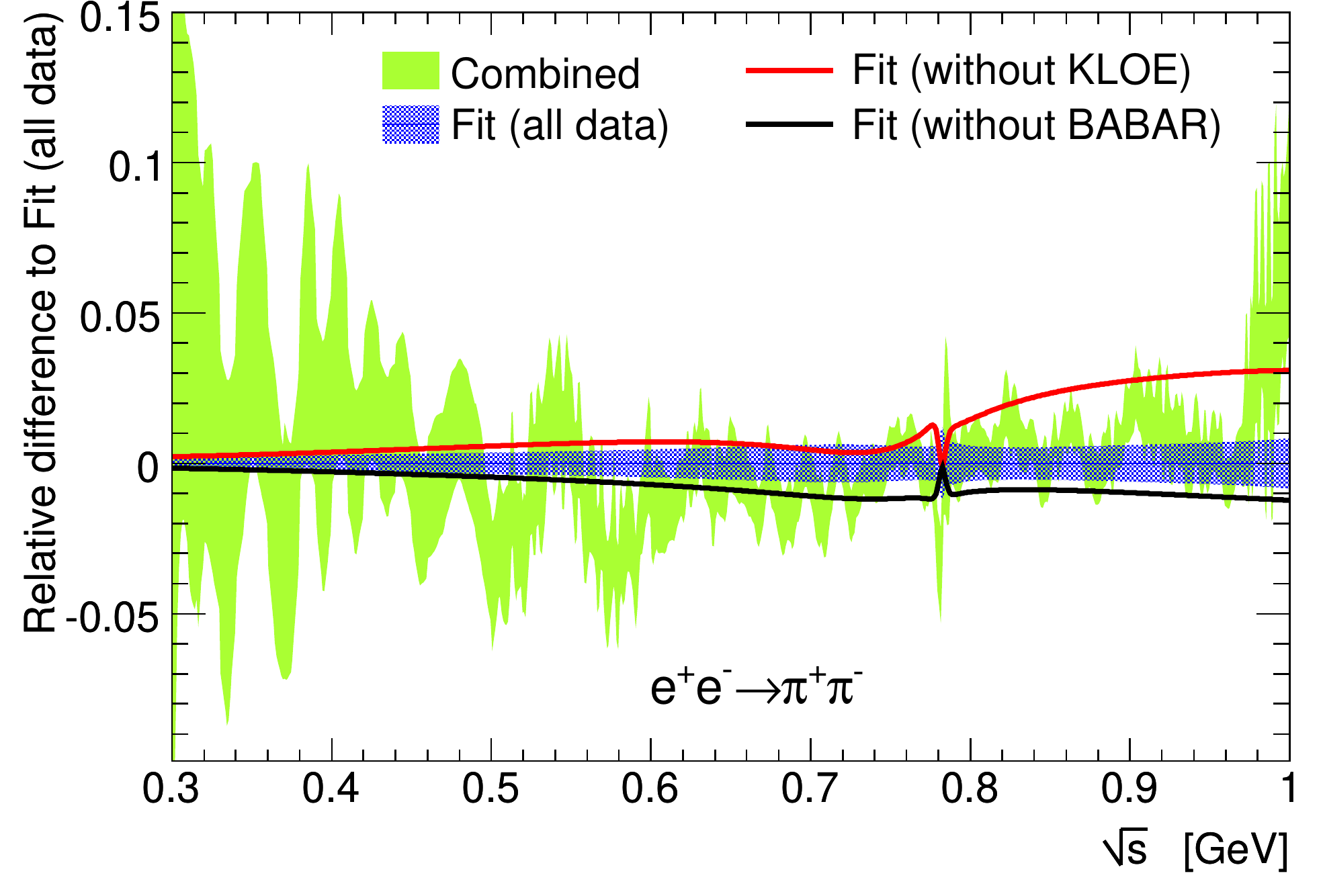}
\end{center}
  \vspace{-0.3cm}
  \caption{The HVPTools combination (green band) relative to the result of the fit to all individual $\pp$ data (blue band)  versus centre-of-mass energy. The black and red curves show the results of two alternative fits where the data from KLOE and BABAR, respectively,   were excluded.}
  \label{fig:ff2_comp}
\end{figure}
The evaluation of the complete \amuhadLOpp integral for the \pp contribution from threshold to 1.8$\;$GeV, using the fit up to 0.6$\;$GeV and the HVPTools data combination above, gives $507.0\pm1.9$. The choice of the ranges is justified by the good agreement between fit and combined data integration in the 0.6--1.0$\;$GeV region with, however, no advantage in precision for the fit. The correlation among the two contributions is found to be 62\% using pseudo-experiments.

Removing BABAR or KLOE from the dataset gives $505.1\pm2.1$ and $510.6\pm2.2$, respectively, with an absolute difference of 5.5 that is significantly larger than the individual uncertainties. Figure~\ref{fig:amu2pi} shows a comparison among the most precise \amuhadLOpp evaluations in the interval 0.6--0.9$\;$GeV. The results of all other experiments fall in-between the BABAR and KLOE results, with insufficient precision to resolve the discrepancy. 

Figure~\ref{fig:ff2_comp} compares the HVPTools combination and the fits without using the BABAR  and KLOE data, respectively, with the fit result for the full dataset. In light of this discrepancy, which is not fully captured by the local uncertainty rescaling procedure, we add as additional systematic uncertainty half of the full difference between the complete integrals without BABAR and KLOE, respectively, and we place the central value of the \amuhadLOpp contribution half-way between the two results. To avoid double counting, the local uncertainty rescaling between BABAR and KLOE is not applied, but that between these and the other \pp datasets is kept.
This procedure results in a total \pp contribution of  $\amuhadLOpp=507.9\pm0.8\pm3.2$, where the first uncertainty is statistical and the second systematic (dominated by the new uncertainty of 2.8).

\subsection{The \sansmath$\mathbf{\boldsymbol{K^+K^-}}$ channel}

\begin{figure}[p]
\includegraphics[width=\figsize]{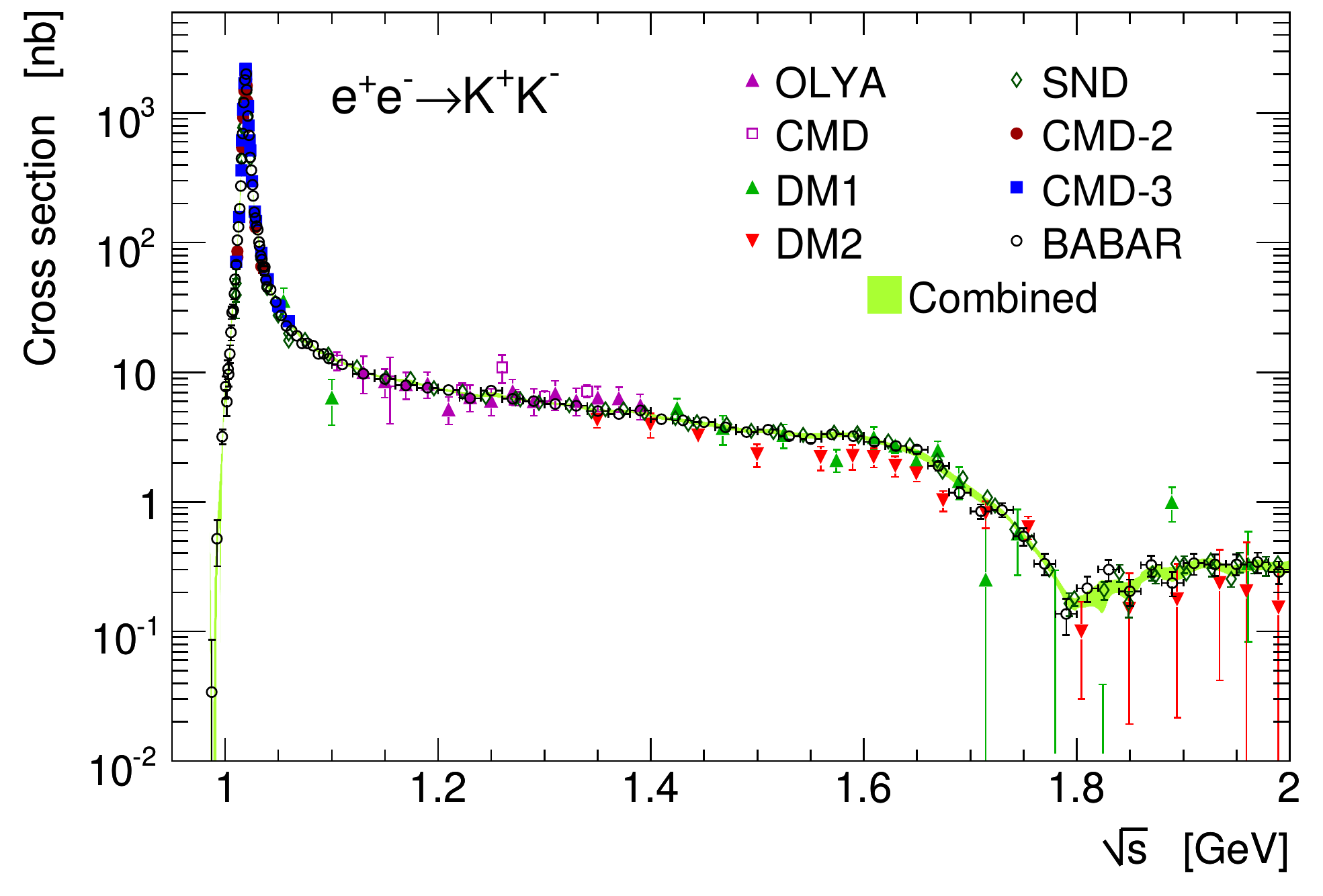}
  \vspace{0.1cm}

\includegraphics[width=\figsize]{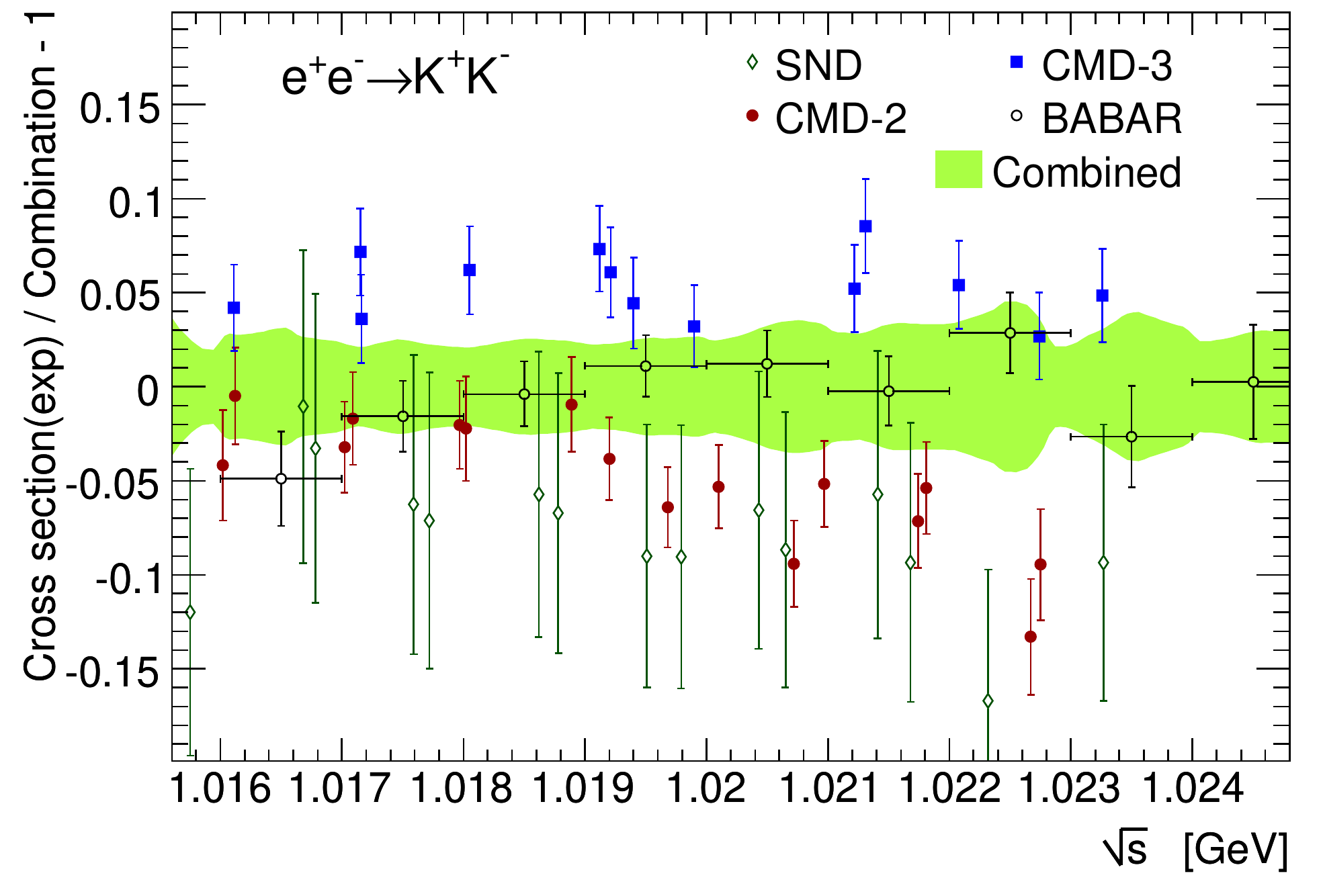}\hspace{\fighspace}

  \vspace{0.1cm}
\includegraphics[width=\figsize]{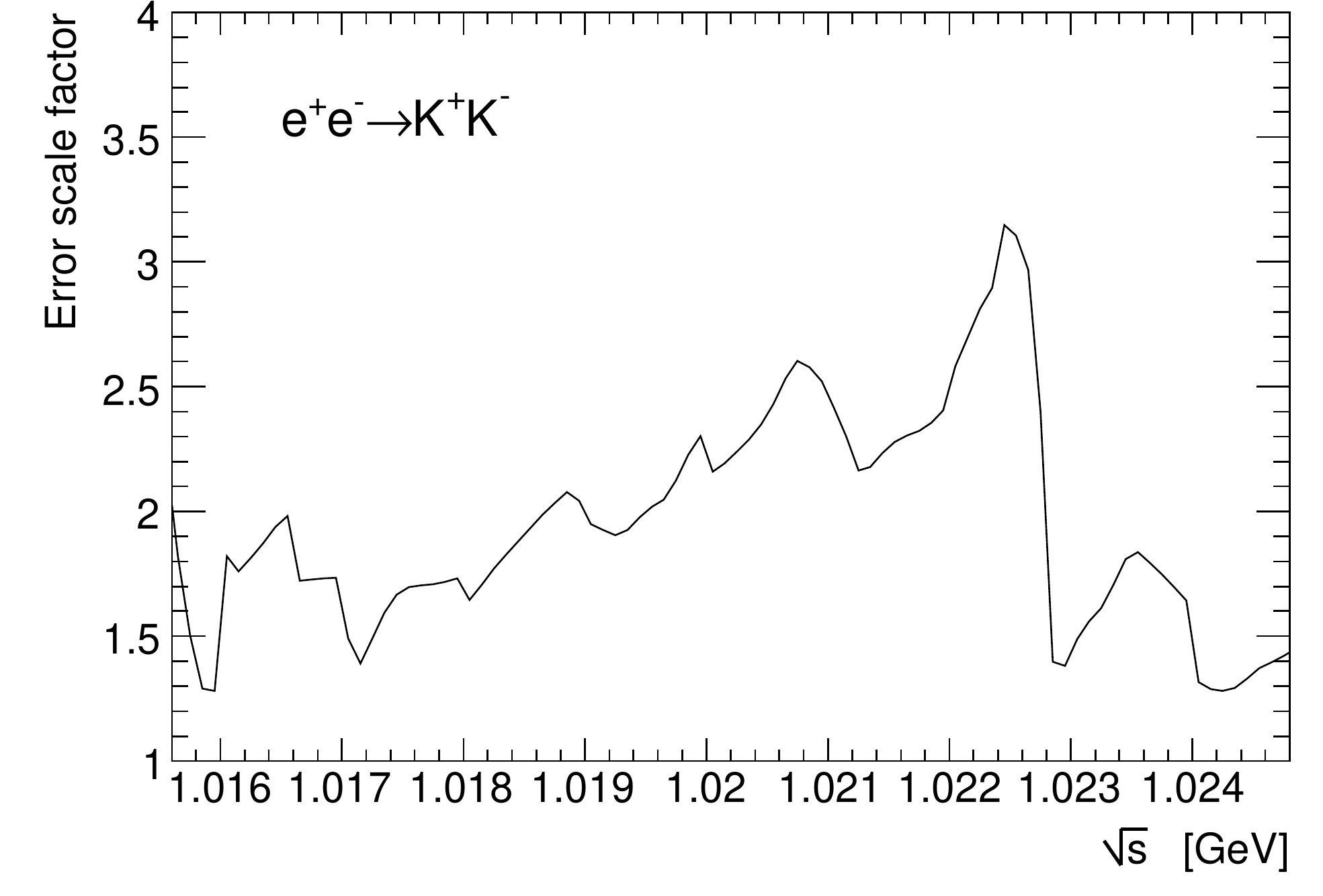}
\vspace{-0.1cm}

\caption[.]{ 
            Top panel: bare cross sections for  $e^+e^-\to K^+ K^-  $. See text for a description of the data used. 
            Middle: comparison between individual $\ee\to K^+K^-$ cross-section measurements from BABAR~\cite{babar-kpkm}, CMD-2~\cite{Akhmetshin:2008gz}, CMD-3~\cite{Kozyrev:2017agm} and SND~\cite{snd-kpkm}, and the HVPTools combination.  
            Bottom: local scale factor versus centre-of-mass energy applied 
            to the combined $K^+K^-$ cross-section uncertainty to account for inconsistency 
            in the individual measurements. }
\label{fig:kkweights}
\end{figure}

Tensions among datasets are also present in the $K^+K^-$ channel (see top panel of Fig.~\ref{fig:kkweights} for a display of the available measurements). A discrepancy between BABAR and SND was observed for masses between 1.05 and 1.4$\;$GeV, which has been resolved with the most recent SND result~\cite{snd-kpkm} so that older SND data are discarded. 

Concerns also arise regarding data on the $\phi$(1020) resonance. Previously, a 5.1\% difference between CMD-2 at VEPP-2M and BABAR was observed, with the CMD-2 data being lower. New results from CMD-3 at VEPP-2000~\cite{Kozyrev:2017agm} exhibit the opposite effect: they are 5.5\% higher than BABAR (cf. middle panel in Fig.~\ref{fig:kkweights}). The  discrepancy of almost 11\% between the  CMD-2 and CMD-3 datasets, which largely exceeds the  quoted systematic uncertainty of 2.2\%, of which only 1.2\% accounts for uncertainties in the detection efficiency, is claimed to originate from a better understanding of the detection efficiency of low-energy kaons in the CMD-3 data.\footnote{In comparison with the CMD-2/3 and SND measurements, the ISR method of BABAR benefits from higher-momentum kaons with better detection efficiency owing to the boost of the final state.}
Given the  yet unresolved situation, we keep both CMD-2 and CMD-3 datasets, which due to the uncertainty rescaling procedure in presence of discrepancies leads to a deterioration of the precision by about a factor of two of the combined data (cf. bottom panel of Fig.~\ref{fig:kkweights}).\footnote{We have verified that the local $\chi^2$ rescaling procedure covers the global discrepancy among the CMD-2 and CMD-3 data by removing alternatively one or the other dataset from the $\Kp\Km$ combination. The difference of 0.45 resulting between the two $a_\mu$ values is covered by the uncertainty rescaling (a similar conclusion is reached for \dahadZ). There is therefore no need to introduce an additional global systematic uncertainty as for the $\pp$ case.}

\subsection{Other channels}

Recent measurements have been included in the data combinations: $\pi^0 \gamma$ from SND~\cite{Achasov:2018ujw}, $\pi^+\pi^-2\pi^0$ from BABAR~\cite{TheBaBar:2017vzo},  $\pi^+\pi^-3\pi^0$ from BABAR~\cite{Lees:2018dnv}, $\eta\pi^+\pi^-$ from BABAR~\cite{TheBABAR:2018vvb}, $\eta\pi^+\pi^-\pi^0$ from CMD-3~\cite{CMD-3:2017tgb} and SND~\cite{Achasov:2019duv}, $\phi\eta$ from CMD-3~\cite{Ivanov:2019crp}, and $K_SK_L\pi^0$ from SND~\cite{Achasov:2017vaq}.
The  $\pi^0 \gamma$ and $\pi^+\pi^-\pi^0$ contributions include small additions of $0.12\pm 0.01$ and $0.01\pm 0.00$, respectively, to cover the threshold region up to the lowest-energy data measurements~\cite{hmnt03,knt19}.

Only very few final states remain to be estimated using isospin symmetry. Already in 2017, a significant step was achieved with the BABAR measurements of all the final states contributing to the $K \Kbar \pi$ and $K \Kbar \pi\pi$ channels, so that previous isospin-based estimates became obsolete. Now the only significant (albeit small) contribution obtained with the use of isospin constraints is that for the $\pi^+\pi^-4\pi^0$ channel. The part excluding $\eta 3\pi$, which is obtained from measured processes and reevaluated in this analysis, amounts to a fraction of the total dispersion integral of only 0.016\% with an assigned systematic uncertainty of 100\%. One could also question the completeness of the set of exclusive processes considered below 1.8 GeV, including up to 6-pion and 
$K \Kbar$+3 pions. A recent measurement of the $3\pi^+ 3\pi^- \pi^0$ cross section by CMD-3~\cite{cmd3-7pi} allows one to estimate a very small total 7-pion contribution, included in this analysis, of only 0.002\%. Although such high-multiplicity channels appear to be contributing negligibly below 1.8$\;$GeV their importance is likely to increase above.

All other contributions are identical to the ones described in our previous analysis~\cite{dhmz2017}, except for (i) a reevaluation of the contribution from $\omega$ decay modes not reconstructed in other exclusive channels, and (ii) a better estimate of the $K \overline{K} \pi^+ \pi^- \pi^0$ contribution, excluding $\phi\eta$ which is dominated by the $K \overline{K} \omega$ final state~\cite{Aubert:2007ef}.

\section{~Compilation and results}
\label{sec:Results}

\begin{table*}[p]
\newcommand{\gam}{\ensuremath{\gamma}\xspace}
\setlength{\tabcolsep}{0.0pc}
\begin{tabularx}{\textwidth}{@{\extracolsep{\fill}}lrr} 
\hline\noalign{\smallskip}
Channel &   \amuhadLO $[10^{-10}]$ & \dahadZ $[10^{-4}]$ \\
\noalign{\smallskip}\hline\noalign{\smallskip}

$\pi^0\gamma$ &$  4.41 \pm 0.06 \pm 0.04 \pm 0.07$&$  0.35 \pm 0.00 \pm 0.00 \pm 0.01$\\
$\eta\gamma$ &$  0.65 \pm 0.02 \pm 0.01 \pm 0.01$&$  0.08 \pm 0.00 \pm 0.00 \pm 0.00$\\
$\pi^+\pi^-$ &$507.85 \pm 0.83 \pm 3.23 \pm 0.55$&$ 34.50 \pm 0.06 \pm 0.20 \pm 0.04$\\
$\pi^+\pi^-\pi^0$ &$ 46.21 \pm 0.40 \pm 1.10 \pm 0.86$&$  4.60 \pm 0.04 \pm 0.11 \pm 0.08$\\
$2\pi^+2\pi^-$ &$ 13.68 \pm 0.03 \pm 0.27 \pm 0.14$&$  3.58 \pm 0.01 \pm 0.07 \pm 0.03$\\
$\pi^+\pi^-2\pi^0$ &$ 18.03 \pm 0.06 \pm 0.48 \pm 0.26$&$  4.45 \pm 0.02 \pm 0.12 \pm 0.07$\\
$2\pi^+2\pi^-\pi^0~(\eta~\textrm{excl.})$ &$  0.69 \pm 0.04 \pm 0.06 \pm 0.03$&$  0.21 \pm 0.01 \pm 0.02 \pm 0.01$\\
$\pi^+\pi^-3\pi^0~(\eta~\textrm{excl.})$ &$  0.49 \pm 0.03 \pm 0.09 \pm 0.00$&$  0.15 \pm 0.01 \pm 0.03 \pm 0.00$\\
$3\pi^+3\pi^-$ &$  0.11 \pm 0.00 \pm 0.01 \pm 0.00$&$  0.04 \pm 0.00 \pm 0.00 \pm 0.00$\\
$2\pi^+2\pi^-2\pi^0~(\eta~\textrm{excl.})$ &$  0.71 \pm 0.06 \pm 0.07 \pm 0.14$&$  0.25 \pm 0.02 \pm 0.02 \pm 0.05$\\
$\pi^+\pi^-4\pi^0~(\eta~\textrm{excl., isospin})$ &$  0.08 \pm 0.01 \pm 0.08 \pm 0.00$&$  0.03 \pm 0.00 \pm 0.03 \pm 0.00$\\
$\eta\pi^+\pi^-$ &$  1.19 \pm 0.02 \pm 0.04 \pm 0.02$&$  0.35 \pm 0.01 \pm 0.01 \pm 0.01$\\
$\eta\omega$ &$  0.35 \pm 0.01 \pm 0.02 \pm 0.01$&$  0.11 \pm 0.00 \pm 0.01 \pm 0.00$\\
$\eta \pi^+\pi^-\pi^0 (\textrm{non-}\omega,\phi)$ &$  0.34 \pm 0.03 \pm 0.03 \pm 0.04$&$  0.12 \pm 0.01 \pm 0.01 \pm 0.01$\\
$\eta2\pi^+2\pi^-$ &$  0.02 \pm 0.01 \pm 0.00 \pm 0.00$&$  0.01 \pm 0.00 \pm 0.00 \pm 0.00$\\
$\omega\eta\pi^0$ &$  0.06 \pm 0.01 \pm 0.01 \pm 0.00$&$  0.02 \pm 0.00 \pm 0.00 \pm 0.00$\\
$\omega\pi^0~(\omega\rightarrow\pi^0\gamma)$ &$  0.94 \pm 0.01 \pm 0.03 \pm 0.00$&$  0.20 \pm 0.00 \pm 0.01 \pm 0.00$\\
$\omega2\pi~(\omega\rightarrow\pi^0\gamma)$ &$  0.07 \pm 0.00 \pm 0.00 \pm 0.00$&$  0.02 \pm 0.00 \pm 0.00 \pm 0.00$\\
$\omega~(\textrm{non-}3\pi,\pi\gamma,\eta\gamma)$ &$  0.04 \pm 0.00 \pm 0.00 \pm 0.00$&$  0.00 \pm 0.00 \pm 0.00 \pm 0.00$\\
$K^+K^-$ &$ 23.08 \pm 0.20 \pm 0.33 \pm 0.21$&$  3.35 \pm 0.03 \pm 0.05 \pm 0.03$\\
$K_SK_L$ &$ 12.82 \pm 0.06 \pm 0.18 \pm 0.15$&$  1.74 \pm 0.01 \pm 0.03 \pm 0.02$\\
$\phi~(\textrm{non-}K\ol{K},3\pi,\pi\gamma,\eta\gamma)$ &$  0.05 \pm 0.00 \pm 0.00 \pm 0.00$&$  0.01 \pm 0.00 \pm 0.00 \pm 0.00$\\
$K\overline{K}\pi$ &$  2.45 \pm 0.05 \pm 0.10 \pm 0.06$&$ 0.78 \pm 0.02 \pm 0.03 \pm 0.02$\\
$K\overline{K}2\pi$ &$  0.85 \pm 0.02 \pm 0.05 \pm 0.01$&$ 0.30 \pm 0.01 \pm 0.02 \pm 0.00$\\
$K\overline{K}\omega$ &$ 0.00 \pm 0.00 \pm 0.00 \pm 0.00$&$ 0.00 \pm 0.00 \pm 0.00 \pm 0.00$\\
$\eta\phi$ &$  0.33 \pm 0.01 \pm 0.01 \pm 0.00$&$  0.11 \pm 0.00 \pm 0.00 \pm 0.00$\\
$\eta K\ol{K}~(\textrm{non-}\phi)$ &$  0.01 \pm 0.01 \pm 0.01 \pm 0.00$&$  0.00 \pm 0.00 \pm 0.01 \pm 0.00$\\
$\omega3\pi~(\omega\rightarrow\pi^0\gamma)$ &$  0.06 \pm 0.01 \pm 0.01 \pm 0.01$&$  0.02 \pm 0.00 \pm 0.00 \pm 0.00$\\
$7\pi~(3\pi^+3\pi^-\pi^0+\textrm{estimate})$ &$  0.02 \pm 0.00 \pm 0.01 \pm 0.00$&$  0.01 \pm 0.00 \pm 0.00 \pm 0.00$\\
\noalign{\smallskip}\hline\noalign{\smallskip}
$J/\psi$ (BW integral) &$  6.20 \pm 0.11$&$  7.00 \pm 0.13$\\
$\psi(2S)$ (BW integral) &$  1.56 \pm 0.05$&$  2.48 \pm 0.08$\\
\noalign{\smallskip}\hline\noalign{\smallskip}
$R~\textrm{data}\,[3.7-5.0]$ GeV &$  7.29 \pm 0.05 \pm 0.30 \pm 0.00$&$ 15.79 \pm 0.12 \pm 0.66 \pm 0.00$\\
\noalign{\smallskip}\hline\noalign{\smallskip}
$R_\textrm{QCD}\,[1.8-3.7~\textrm{\rm GeV}]_{uds}$ &$ 33.45 \pm 0.28 \pm 0.65_\textrm{dual}$&$ 24.27 \pm 0.18 \pm 0.28_\textrm{dual}$\\
$R_\textrm{QCD}\,[5.0-9.3~\textrm{GeV}]_{udsc}$ &$  6.86 \pm 0.04$&$ 34.89 \pm 0.18$\\
$R_\textrm{QCD}\,[9.3-12.0~\textrm{GeV}]_{udscb}$ &$  1.20 \pm 0.01$&$ 15.53 \pm 0.04$\\
$R_\textrm{QCD}\,[12.0-40.0~\textrm{GeV}]_{udscb}$ &$  1.64 \pm 0.00$&$ 77.94 \pm 0.13$\\
$R_\textrm{QCD}\,[>40.0~\textrm{GeV}]_{udscb}$ &$  0.16 \pm 0.00$&$ 42.70 \pm 0.05$\\
$R_\textrm{QCD}\,[>40.0~\textrm{GeV}]_t$ &$  0.00 \pm 0.00$&$ -0.72 \pm 0.01$\\
\noalign{\smallskip}\hline\noalign{\smallskip}
\textbf{Sum}                                                    &$694.0 \pm 1.0 \pm 3.5 \pm 1.6 \pm 0.1_\psi \pm 0.7_\textrm{QCD}$&$275.29 \pm 0.15 \pm 0.72 \pm 0.23 \pm 0.15_\psi  \pm 0.55_\textrm{QCD}$\\
\noalign{\smallskip}\hline
\vspace{-0.2cm}
\end{tabularx}
  \caption[.]{\label{tab:results}
    Compilation of the  contributions to \amuhadLO and \dahadZ  as obtained from HVPTools, and the phenomenological fit for the \pp contribution below 0.6$\;$GeV. Where three (or more) uncertainties are given, the first is statistical, the second channel-specific systematic, and the third common systematic, which is correlated with at least one other channel. For the contributions computed from QCD, only total uncertainties are given, which include effects from the $\as$ uncertainty, the truncation of the perturbative series at four loops, the FOPT vs.\ CIPT ambiguity, and quark mass uncertainties. Except for the latter uncertainty, all other uncertainties are taken to be fully correlated among the various energy regions where QCD is used. The additional uncertainty dubbed ``dual" estimates possible quark-hadron duality violating effects in the QCD estimate between 1.8 and 2.0$\;$GeV. The uncertainties in the Breit-Wigner integrals of the narrow resonances $J/\psi$ and $\psi(2S)$ are dominated by the the respective electronic width measurements~\cite{pdg}. The uncertainties in the sums (last line) are obtained by quadratically adding all statistical and channel-specific systematic uncertainties, and by linearly adding correlated inter-channel systematic uncertainties. 
}
\end{table*}

A compilation of the various contributions to \amuhadLO and to \dahadZ, as well as the total 
results are given in Table~\ref{tab:results}. The experimental uncertainties are separated
into statistical, channel-specific systematic, and common systematic contributions
that are correlated with at least one other channel. 
\begin{figure*}[t]
  \centering
  \includegraphics[width=12.5cm]{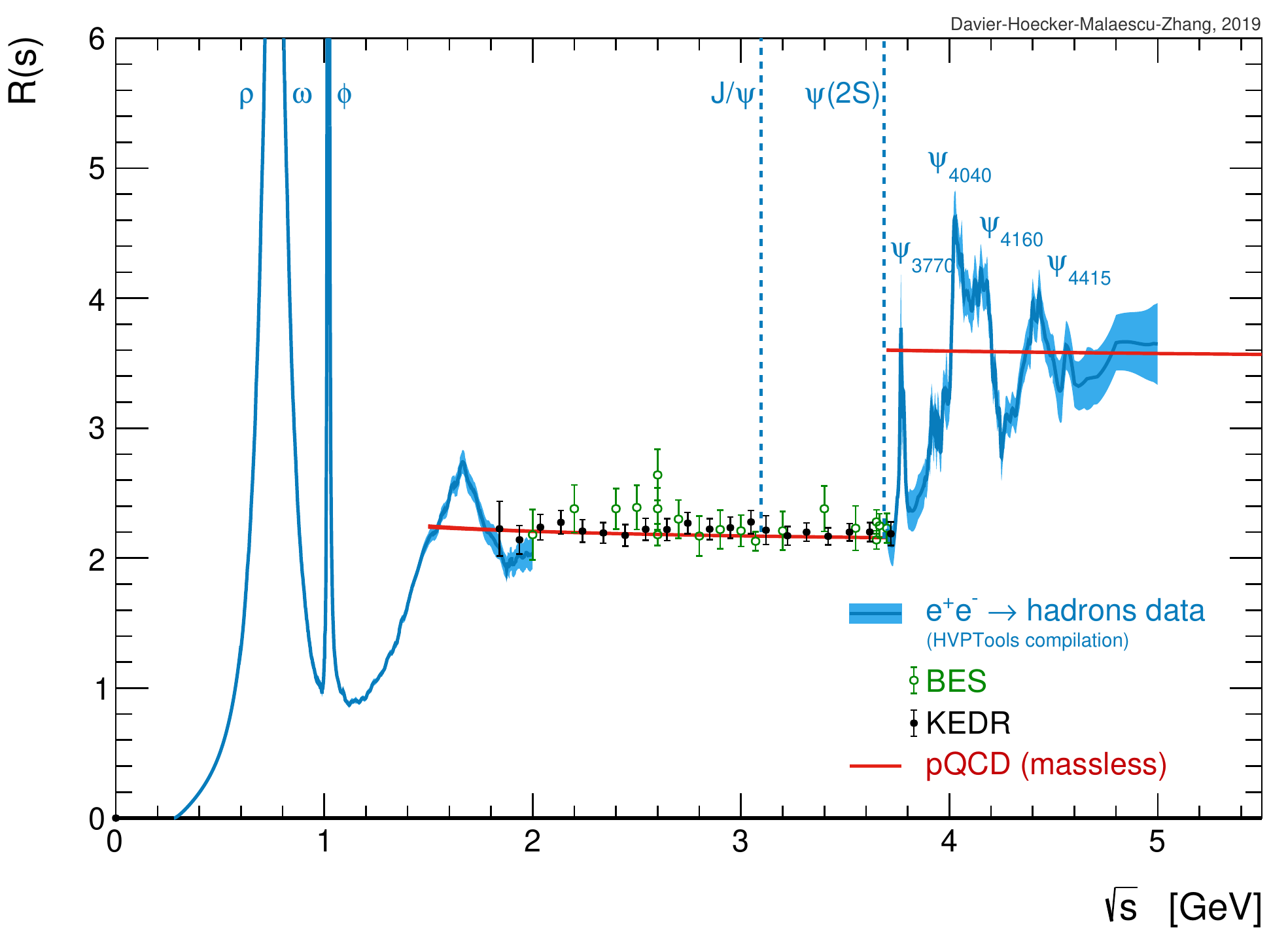}
  \vspace{0.2cm}
  \caption{
             The total hadronic \ee annihilation rate $R$ as a function of centre-of-mass energy. Inclusive measurements from BES~\cite{besR}  
             and KEDR~\cite{kedr-r-1,kedr-r-2} are shown as data points,
             while the sum of exclusive channels from this analysis is given by the narrow blue bands. Also shown for the purpose of illustration is the prediction from massless perturbative QCD (solid red line). 
}
  \label{fig:R}
\end{figure*}
The contributions from the $J/\psi$ and $\psi(2S)$ resonances in Table~\ref{tab:results} are obtained by numerically integrating the corresponding undressed\footnote{The undressing uses the BABAR programme {\tt AFKVAC}, correcting for both leptonic  and hadronic VP effects. The hadronic part is obtained from a numerical integration over cross section data for the continuum, supplemented by analytical expressions for the contributions of narrow resonances including both their real and imaginary components. The resulting correction factors reduce the $J/\psi$ and $\psi(2S)$ contributions to \amuhadLO by about 4\% and are known to a precision of better than $10^{-3}$.} Breit-Wigner lineshapes. The uncertainties in the integrals are dominated by the knowledge of the corresponding electronic width $\Gamma_{R\to ee}$ 
for which we use the values $5.53 \pm 0.10\;$keV for $R=J/\psi$ and $2.34 \pm 0.04\;$keV for $R=\psi(2S)$~\cite{pdg}.

Sufficiently far from the quark thresholds we use four-loop~\cite{baikov} perturbative QCD, including ${\cal O}(\as^2)$ quark mass corrections~\cite{kuhnmass}, to compute the inclusive hadronic cross section. Nonperturbative contributions at $1.8\:\gev$ were determined from data~\cite{dh98} and found to be small. The uncertainties of the $R_{\rm QCD}$ contributions given in Table~\ref{tab:results} are obtained from the quadratic sum of the uncertainty in $\as$  (we use $\asZ=0.1193\pm0.0028$ from the fit to  $Z$ precision data~\cite{gfitter}), the truncation of the perturbative series (we use the full four-loop contribution as systematic uncertainty), the  difference between fixed-order perturbation theory  (FOPT) and, so-called, contour-improved perturbation theory (CIPT)~\cite{ledibpich}, as well as quark mass uncertainties (we use the values and uncertainties from Ref.~\cite{pdg}). The former three uncertainties are taken to be fully correlated between the various energy regions (see Table~\ref{tab:results}), whereas the (smaller) quark-mass uncertainties are taken to be uncorrelated. 

To examine the transition region between the sum of exclusive measurements and QCD we have computed \amuhadLO and \dahadZ in the narrow energy interval $1.8$--$2.0\;\gev$. For the former quantity we find $7.65  \pm 0.31$ and $8.30 \pm 0.09$ for data and QCD, respectively. The full difference of $0.65$ ($0.28\cdot10^{-4}$ in the case of \dahadZ) is assigned as additional systematic uncertainty, labelled by ``dual" subscripts in Table~\ref{tab:results}. It accounts for possible low-mass quark-hadron duality violation effects in the perturbative QCD approximation that we use for this interval to avoid systematic effects due to  unmeasured high-multiplicity channels. 

Figure~\ref{fig:R} shows the total hadronic \ee annihilation rate $R$ versus centre-of-mass energy as obtained from the sum of exclusive data below 2$\;$GeV and from inclusive data between 1.8 and 5$\;$GeV.\footnote{We have verified that the integration of the finely binned $R$ distribution shown in Fig.~\ref{fig:R}, together with its covariance matrix, accurately reproduces the \amuhadLO and \dahadZ results obtained by summing the exclusive modes below 1.8$\;$GeV  in Table~\ref{tab:results}.} Also indicated are the perturbative QCD prediction above 1.5$\;$GeV and the analytical narrow $J/\psi$ and $\psi(2S)$ resonances.

\vspace{0.0cm}
\paragraph*{\bf\em Muon magnetic anomaly\\[0.2cm] } 

Adding all lowest-order hadronic contributions together gives
\beq
\label{eq:amuhadlo}
   \amuhadLO = 694.0 \pm 4.0\,,
\eeq
which is dominated by experimental systematic uncertainties (\cf Table~\ref{tab:results} for a separation of the total uncertainty into its components), with an uncertainty of 2.8 originating from the BABAR versus KLOE discrepancy in the \pp channel. The new result is $0.9$ units larger than  our previous evaluation~\cite{dhmz2017}, $693.1\pm3.4$, mostly because we symmetrised the new BABAR/KLOE systematic uncertainty. The total uncertainty is increased by 18\%.  The result without the additional BABAR/KLOE systematic uncertainty is $693.1\pm3.2$.

Adding to~(\ref{eq:amuhadlo}) the contributions from higher order hadronic loops, $-9.87 \pm 0.09$ (NLO) and $1.24\pm0.01$ (NNLO)~\cite{amu-hadnlo}, hadronic light-by-light scattering, $10.5\pm 2.6$~\cite{amu-lbl},  as well as QED, $11\,658\,471.895 \pm 0.008$~\cite{amu-qed} (see also~\cite{pdgg-2rev} and references therein), and electroweak effects, $15.36 \pm 0.10$~\cite{amu-ew},\footnote{When adjusting~\cite{wjm}  the new full 2-loop calculation in Ref.~\cite{amu-ew-num} to physical quark masses, it reproduces the value obtained in~\cite{amu-ew}.} we obtain the complete SM prediction
\beq
\label{eq:amusm}
  \amuSM = 11\,659\,183.1 \pm 4.0 \pm 2.6 \pm 0.1~(4.8_{\rm tot})\,,
\eeq
where the uncertainties account for lowest and higher order hadronic, and 
other contributions, respectively. The result~(\ref{eq:amusm}) deviates from the 
experimental value, $\amuExp=11\,659\,209.1 \pm 5.4 \pm 3.3$~\cite{bnl,pdgg-2rev}, 
by $26.0 \pm 7.9$ ($3.3\sigma$).

\begin{figure}[t]
\vspace{0.1cm}

\includegraphics[width=\columnwidth]{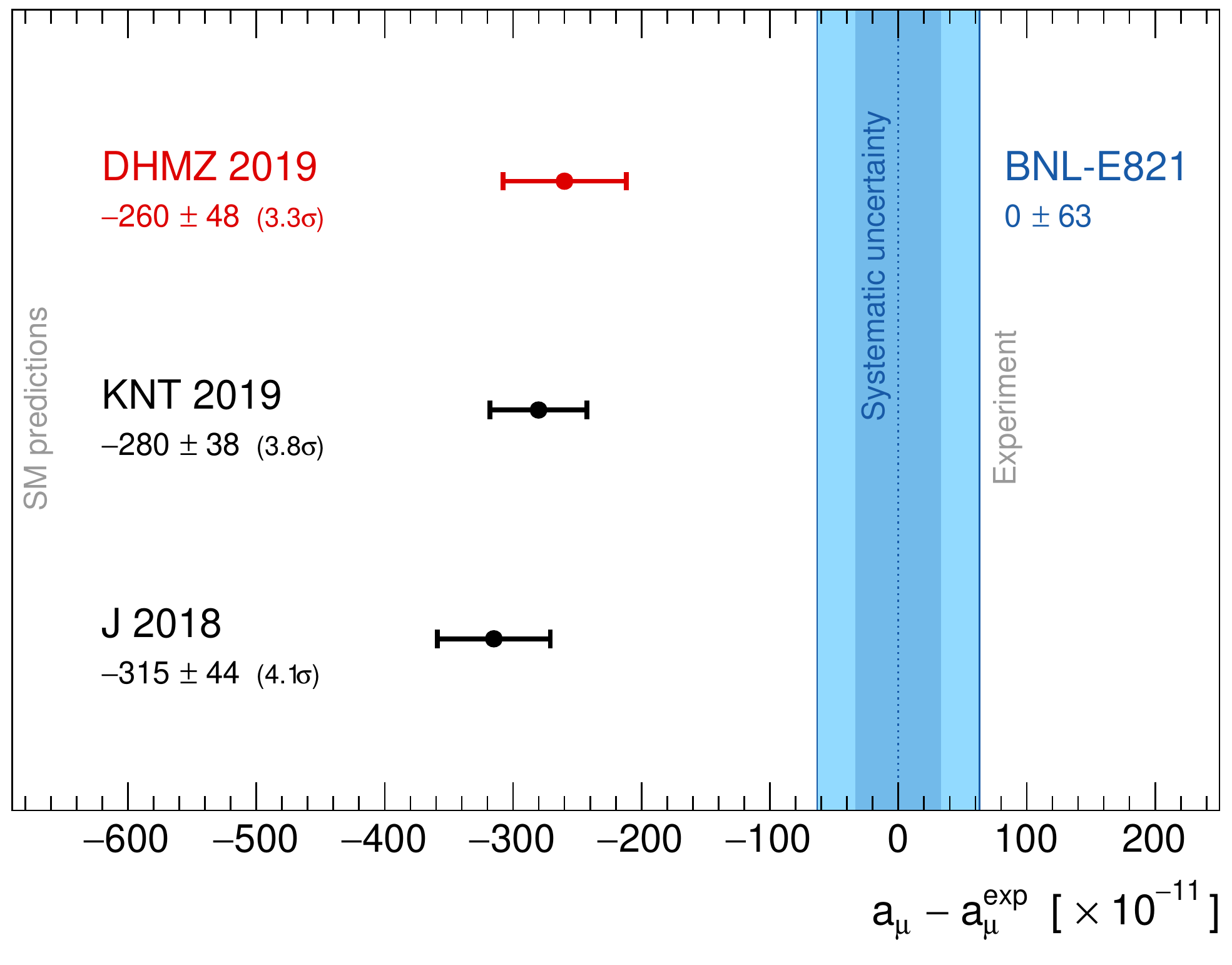}
\vspace{-0.3cm}
\caption{ 
        Compilation of recent data-driven results for $\amuSM$ (in units of $10^{-10}$), subtracted by the central value of the experimental average~\cite{bnl,pdgg-2rev}.  The blue vertical band indicates the experimental uncertainty, with the darker inlet representing the experimental systematic uncertainty. The representative SM predictions are taken from KNT 2019~\cite{knt19}, J 2018~\cite{jeger},  and this work (DHMZ 2019). }
\label{fig:amures}
\end{figure}

A compilation of recent SM predictions for \amu compared with the experimental
result is given in Fig.~\ref{fig:amures}.

\vspace{0.0cm}
\paragraph*{\bf\em Running electromagnetic coupling at \boldmath$m_Z^2$ \\[0.2cm]}

The sum of all quark-flavour terms from Table~\ref{tab:results} gives for the 
hadronic contribution to the running of \aZ
\beq
\label{eq:dahad}
   \dahadZ   = (275.3 \pm 1.0)\cdot 10^{-4}\,,
\eeq
the uncertainty of which is dominated by data systematic effects ($0.7\cdot 10^{-4}$) and the uncertainty in the QCD prediction ($0.6\cdot 10^{-4}$). The use of the same inputs with different integration kernels in the calculations induces a correlation of $+44$\% between the $a_\mu^\mathrm{had, LO}$ and $\Delta \alpha_\mathrm{had}(m^2_Z)$ uncertainties. The result without the new BABAR/KLOE systematic uncertainty is $275.2\pm0.9$.

Adding to~(\ref{eq:dahad}) the four-loop leptonic contribution, $\Delta\alpha_{\rm lep} (m_Z^2)=(314.979 \pm 0.002)\cdot 10^{-4}$~\cite{sturm}, 
one finds
\beq
   \alpha^{-1}(m_Z^2) = 128.947 \pm 0.013\,.
\eeq
The current uncertainty on $ \alpha(m_Z^2)$ is sub-dominant in the SM prediction of the $W$-boson mass (the dominant uncertainties are due to the top mass and of theoretical origin), but dominates the prediction of $\sin^2\theta_{\rm eff}^\ell$, which, however, is about twice more accurate than the combination of all present measurements~\cite{gfitter}.

\section{~Conclusions and perspectives}

Using newest available $e^+e^-\to {\rm hadrons}$ cross-section data we have reevaluated the lowest-order hadronic vacuum polarisation contribution to the Standard Model prediction of the anomalous magnetic moment of the muon, and the hadronic contribution to the running electromagnetic coupling strength at the $Z$-boson mass. For the former quantity we find $\amuhadLO = (694.0 \pm 4.0)\cdot 10^{-10}$. In spite of new data and the use of a more precise fit to evaluate the threshold region up to 0.6$\;$GeV, the uncertainty on this contribution has increased to 0.6\% since our last evaluation~\cite{dhmz2017}, due to the addition of a new systematic uncertainty to account for a global discrepancy between \pp data from BABAR and KLOE.
Resolving this discrepancy would allow to reduce the \amuhadLO uncertainty by 20\%.\footnote{The contribution of the $\pp$ channel to the total \amuhadLO uncertainty-squared is $71\%$.} 

\sloppy
The discrepancy between measurement and complete Standard Model prediction remains at a non-conclusive $3.3\sigma$ level. The new Fermilab $g-2$ experiment currently in operation~\cite{fnal-g-2} aims at up to four times better ultimate precision and has the potential to clarify the situation. 

To match the precision of the new experiment  further progress is  needed to reduce the uncertainty on \amuhadLO from dispersion relations. New analyses of the dominant $\pip\pim$ channel are underway at the BABAR, CMD-3 and SND experiments for which a systematic uncertainty below 0.5\% may be reachable.  It is also important to  improve the precision of the $\pip\pim\piz$ and $K^+ K^-$ channels. The new Belle-2 experiment at the KEK Super-B factory will also contribute to measuring hadronic cross sections via the ISR method once the detector performance is fully understood and sufficient statistics has been accumulated. 

Independently of the data-driven approach, lattice QCD calculations of \amuhadLO  are  also progressing albeit not yet reaching competitive  precision~\cite{Lattice-amu}.

The determination of \amuhadLO is closing in on the estimated  uncertainty of the hadronic light-by-light scattering contribution \amuhadLBL of  $2.6\cdot10^{-10}$, which appears irreducible at present. Here only phenomenological models have been used so far and lattice QCD calculations could have a strong impact~\cite{Lattice-lbl}, as well as a new promising dispersive approach~\cite{Hoferichter:2018kwz}.

%
%


%
%

\end{document}